\documentclass[twocolumn,trackchanges]{aastex631}
\usepackage{amsmath}
\usepackage{newtxtext,newtxmath}
\usepackage[english]{babel}
\usepackage[autostyle]{csquotes}
\usepackage{nicefrac}
\usepackage{float}

\begin{document}

\title{A rare, strong shock front in the merging cluster SPT-CLJ 2031-4037}

\author[0000-0003-3621-2999]{Purva Diwanji}
\affiliation{Department of Physics and Astronomy, The University of Alabama in Huntsville \\
301 Sparkman Drive NW, Huntsville, AL 35899, USA}

\author{Stephen A. Walker}
\affiliation{Department of Physics and Astronomy, The University of Alabama in Huntsville \\
301 Sparkman Drive NW, Huntsville, AL 35899, USA}


\author{Mohammad S. Mirakhor}
\affiliation{Department of Physics and Astronomy, The University of Alabama in Huntsville \\
301 Sparkman Drive NW, Huntsville, AL 35899, USA}






\begin{abstract}
 We present our findings from the new deep \textit{Chandra} observations ($256$ ks) of the merging galaxy cluster SPT-CLJ 2031-4037 at $z = 0.34$. Our observations reveal intricate structures seen in a major merger akin to the Bullet Cluster. The X-ray data confirm the existence of two shock fronts, one to the northwest and one to the southeast by directly measuring the temperature jump of gas across the surface brightness edges. The stronger shock front in the northwest has a density jump of $3.16 \pm 0.34$ across the sharp surface brightness edge and  Mach number  $M = 3.36^{+0.87}_{-0.48}$, which makes this cluster one of the rare merging systems with a Mach number $M > 2$. We use the northwestern shock to compare two models for shock heating - the instant heating model and the Coulomb collisional heating model, and we determine that the temperatures across the shock front agree with the Coulomb collisional model of heating. For the shock front in the southeastern region, we find a density jump of $1.53 \pm 0.14$ and a Mach number of $M = 1.36 ^{+0.09}_{-0.08}$.   

\end{abstract}



\section{Introduction} \label{sec:intro}
Galaxy clusters, the most massive gravitationally bound structures in the universe are formed through hierarchical mergers of smaller subclusters \citep{Dasadia_2016}.  Mergers of galaxy clusters are the most energetic events in the Universe after the Big Bang wherein the sub-clusters collide at velocities of $\sim$10$^3$ km s$^{-1}$, releasing energy of the order of 10$^{64}$ ergs \citep{Sarazin_2002_chapter}.\textbf{  }A fraction of the kinetic energy released during mergers is dissipated into the ICM via shocks and turbulence and may also cause non-thermal phenomena such as amplification of magnetic fields in the ICM, and acceleration of ultrarelativistic particles in the cluster\citep{Sarazin_2008, Blanford_Eichler_1987}. 

Shock fronts, seen as sharp discontinuities in X-ray brightness and temperature, provide a rare chance to observe and investigate such merger systems and their geometry. They are also used to measure gas bulk velocities and to understand transport processes in the ICM, including electron-ion equilibration and thermal conduction, magnetic fields, and turbulence \citep{MV_review_2007, Takizawa_1999}.

SPT-CLJ 2031-4037 (hereafter SPT J2031) is a massive merger system  with $M_{500}\sim 8 \times 10^{14}$ M$_{\odot}$ \citep{spt_mass}, and X-ray luminosity $L_{[0.1-2.4 keV]} = 1.04 \times 10^{45}$ erg s$^{-1}$ \citep{spt_luminosity} at redshift  $z=0.34$ \citep{ROSAT_Bohringer_2004}. The morphologically disturbed cluster \citep{Nurgaliev_2017} was first discovered in a ROSAT-ESO Flux Limited X-ray (REFLEX) Galaxy Cluster survey \citep{ROSAT_Bohringer_2004} as RXCJ2031.8-4037. It was also catalogued via the Sunyaev-Zel'dovich effect by the South Pole Telescope (SPT) \citep{Plagge_2010, Williamson_2011} and the Planck Satellite \citep{Planck_collab}. The redshift of this system is similar to that of the Bullet Cluster (redshift $z=0.3$), and hence the size of their angular features are comparable.  

Previous $10$ ks \textit{Chandra} observation revealed two surface brightness peaks indicating that it is very likely a major merger. Recent radio observations of SPT J2031 performed with the GMRT at $325$ MHz and with VLA (L-band observation) at $1.7$ GHz revealed diffuse radio emission in the cluster \citep{Raja_2020} leading to the speculation of a merger event in the past, which can be confirmed with deep X-ray observations. To investigate the possible occurrence of a shock front, we obtained deep \textit{Chandra} observations. 


In this paper, we present our results from deep \textit{Chandra} observations of SPT J2031, which include the detection of a strong merger shock, the spatially resolved temperature map, and the preferred method of shock-heating. In section \ref{chandra} we outline details of the observations and discuss the data reduction. In sections \ref{image_analysis} and \ref{spatial}  we present the image analysis and show the emissivity, temperature map, pseudo-pressure map, and the GGM filtered image. In section \ref{shock}, we analyze the primary shock and the southeastern edge in more detail by obtaining the surface brightness profiles and temperature profiles. Understanding the process of electron-ion equilibration is crucial in deciphering the complex dynamics of shock fronts in merging galaxy clusters. In this paper, we present a comprehensive analysis of the electron-ion equilibration test performed on the shock fronts observed in the galaxy cluster SPT J2031. The investigation involves the comparison of two prominent models of shock heating: the adiabatic-collisional model and the instant shock-heating model. The post-shock electron temperature profiles are compared to the Coulomb collisional and instant shock heating models for electron-ion equilibration.

We assume a flat cosmology with $H_0 = 70$ km s$^{-1}$ Mpc$^{-1}$, $\Omega_m = 0.3$ and $\Omega_\Lambda = 0.7$. The redshift is $z = 0.34$  where $1"$ corresponds to $4.892$ kpc. All the error bars are at a $68 \% $ confidence level unless stated otherwise.

\section{Chandra Data Analysis} \label{chandra}
SPT J2031 was observed by the  \textit{Chandra} Advanced CCD Imaging Spectrometer (ACIS) detector in the Very Faint (VFAINT) mode for a total of $256$ ks spread over 10 observations (PI: S. A. Walker). All observations were done with the ACIS-S. The obsID, dates of observation, approximate exposure time and cleaned exposure time are listed in Table \ref{SPT_obs_details}.  
 
\begin{table*}
\caption{Details of the deep ($\sim250 ks$) \textit{Chandra} observations of the SPT J2031 Cluster utilised for the analysis shown in this paper. }
 \centering 
     \begin{tabular}{|llllll|} 
 \hline
Obs ID & RA & Dec  & Date & Exp time (ks)  & Cleaned time (ks) \\ [0.5ex] 
 \hline
21539  & 20 31 51.10 & -40 37 22.10 & 2019 Aug 05 & 36.0 & 32.8  \\ 

24505  & 20 31 51.64 & -40 37 19.64 &  2021 Aug 04 & 29.7 & 27.6  \\ 

24508  &  20 31 51.64 & -40 37 19.64 &  2021 Aug 09 & 27.7 & 26.4   \\ 

24510  &  20 31 51.64 & -40 37 19.64 & 2021 Aug 23 & 22.75 & 20.7 \\

24509  & 20 31 51.64 & -40 37 19.64 &  2021 Aug 28 & 32.6 & 30.6   \\ 

24507  & 20 31 51.64 & -40 37 19.64 &  2021 Nov 28 & 19.8 & 17.7   \\

26215 & 20 31 51.64 & -40 37 19.64 &  2021 Nov 28 & 9.9 & 8.4    \\

24506  & 20 31 51.64 & -40 37 19.64 &  2021 Nov 30 & 24.7 & 22.7   \\

23843  & 20 31 51.64 & -40 37 19.64 &  2022 Jul 26 & 19.8 & 17.8    \\

26479 & 20 31 51.64 & -40 37 19.64 &  2022 Jul 29 & 23.3 & 20.4    \\
 \hline 
\end{tabular}
    \label{SPT_obs_details}
\end{table*}

\subsection{Data Reduction} \label{data_reduction}
All data reduction was performed using CIAO, \textit{Chandra's} data analysis system \citep{CIAO_software} (version 4.14) and CALDB, the calibration database (version 4.10.2) provided by the \textit{Chandra} X-ray Center (CXC).  The primary data set given by the detector is an events list file of photons with measurements like the spatial resolution of the X-ray photons that arrive at the detector, the time of arrival, and the energy of that photon, called the event 1 files. These event 1 files were reprocessed using the \textsl{chandra\_repro} script, taking into account the most recent calibrations to the detector, by applying the latest charge transfer inefficiency (CTI) correction, time-dependent gain adjustment, gain map, to obtain the appropriate response files, new bad pixel files and the processed level 2 event files. 
The \textsl{deflare} routine, which uses the \textsl{lc\_clean} script created by M. Markevitch was used to detect and get rid of flares and periods of anomalously low count rates from the input light curves. As can be seen in table \ref{SPT_obs_details}, the data were mostly clean and the final cleaned exposure was $225$ ks.

The cleaned and reprocessed files were reprojected to create a merged image using \textsl{merge\_obs} in the softband ($0.5-2.5$ keV) and in the broadband ($0.7 - 7.0$keV) and a merged event file and exposure-corrected images were created. The \textsl{merge\_obs} script combines the \textsl{reproject\_obs} and \textsl{flux\_obs} script. The \textsl{reproject\_obs} script finds the appropriate ancillary response files (ARF) for all the event 2 files, matches up with the observations and creates a new single event file by merging the event files of individual observations. The \textsl{flux\_obs} script creates exposure maps and the exposure-corrected image.  The bright point sources in the exposure-corrected image were removed by first excluding the regions by eye, and the excluded regions were filled in using the \textsl{dmfilth} script. This script replaces the pixel values in the excluded regions of the image with values interpolated from the surrounding regions using a Poisson probability distribution. 

Blank-sky observations were extracted using the \textsl{blanksky} script which were then reprojected to match the coordinates of the observation. The blank-sky backgrounds were normalized by matching their count rate in the $9.5-12$ keV energy band to that of the observed dataset, thus ensuring uniformity.

\section{Image Analysis} \label{image_analysis}
\begin{figure*}
    \centering
	\includegraphics[width=0.9\textwidth]{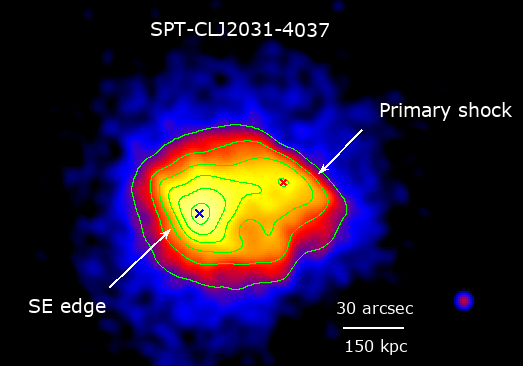}
    \caption{The exposure corrected image of SPT J2031 with the point sources removed in the $0.5 - 7.0$ keV energy range, smoothed with Gaussian $\sigma$ = 3. North is up and East is to the left. Two sharp surface brightness edges are seen here, the Primary Shock in the northwest and the SE edge in the southeast of the image. The brightest X-ray peak lies behind the SE edge, marked by the blue cross. An additional X-ray peak lies behind the primary shock, marked by a red cross. The green lines represent \textit{Chandra} contours.  }
    \label{fig:spt_no_pt_srcs_labelled}
\end{figure*}

Fig. \ref{fig:spt_no_pt_srcs_labelled} shows an exposure-corrected image of the cluster created by combining all the individual \textit{Chandra} observations, with the point sources removed in the $0.5-7.0$ keV energy band. The geometry of the image suggests that the system recently underwent a merger where the sub-clusters passed through each other along the east-west direction.  The X-ray emission is seen extended from the SE to the NW direction. Two sharp surface brightness edges can be seen here, the \enquote{Primary Shock} in the northwestern region and the \enquote{SE edge} in the southeastern region. The brightest X-ray peak lies behind the SE edge and is marked by a blue cross in fig \ref{fig:spt_no_pt_srcs_labelled}. A secondary X-ray peak marked by a red cross in fig \ref{fig:spt_no_pt_srcs_labelled} lies behind the Primary shock in the Northwest. In previous shallow $10$ ks observations, only two bright peaks could be observed, and no edges were visible. These deep \textit{Chandra} observations have helped to resolve the sharp brightness edges and also allow us to produce a more detailed temperature map.



 \begin{figure}
    \centering
    \includegraphics[width=0.9\columnwidth]{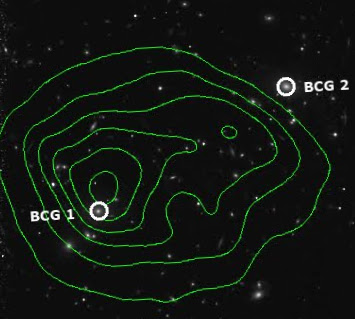}
    \caption{The grayscale image is HST (Hubble Space Telescope) image of SPT J2031 obtained using the F 814W filter. The white dashed circles show the two sub-clusters with their Brightest Cluster Galaxy (BCG). BCG 1 is close to the primary X-ray peak, and BCG 2 is approximately at the location of the primary shock front. The coordinates are shown to be accurate by a number of well-matched point sources.}
    \label{SPT_optical}
\end{figure}

In Fig. \ref{SPT_optical}, \textit{Chandra} X-ray contours from our new observations are superimposed on an HST (Hubble Space Telescope) image of SPT J2031. The grayscale image is the HST image of SPT J2031 obtained by using the F814W filter. The \textit{Chandra} contours are overlaid on this optical image in green. The white dashed circles show the two Brightest Cluster Galaxies (BCG). BCG 1 is close to the primary X-ray peak, and BCG 2 is approximately at the location of the primary shock front and offset from the secondary X-ray peak. The direction of the merger axis is estimated to be roughly from the northwest (NW) to the southeast (SE), passing through the center of the two galaxy distributions.

The BCG 2 shown in the figure is SMACSJ2031.8-4036, which has been extensively studied by deep HST and MUSE as it is a strong lensing cluster. According to the mass modelling presented in \cite{Richard_SMAC}, the eastern component has a mass $M_{east} = 2.4 \times 10^{14}$M$_{\odot}$.

In galaxy cluster mergers, the galaxies within the sub-cluster behave like collisionless particles, and lead the baryonic gas after the collision. This lag between the motion of the sub-cluster galaxies and the baryonic gas can result in an offset between the centroids of the main mass distribution and the elongated peak in the X-ray emission \citep{Canning_2011_BCG}. Comparing the contours representing the brightest X-ray peaks in Fig. \ref{fig:spt_no_pt_srcs_labelled} with the BCGs in Fig. \ref{SPT_optical}, there is an offset of the brightest X-ray peaks from the BCGs, indicating that the system recently underwent a merger. In Fig. \ref{SPT_optical}, the brightest X-ray peak is offset from BCG 1 by $\sim$$0.12$ arcmin ($\sim$ $36$ kpc), and the secondary X-ray peak is offset from BCG 2 by $\sim0.39$ arcmin ($\sim$ $117$ kpc). 
\begin{figure}
    \centering
    \includegraphics[width=0.9\columnwidth]{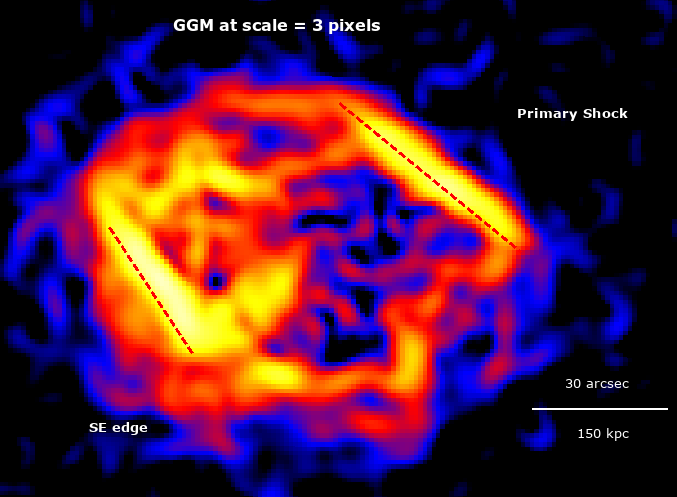}
    \caption{ GGM image of SPT J2031 in the $0.5-7.0$ keV energy range at scale $=3$ pixels.  The red dashed lines highlight the two surfaces with pronounced gradients, the Primary Shock in the northwest and the SE edge in the southeastern direction.}
    \label{fig:ggm}
\end{figure}
We obtained a Gaussian Gradient Magnitude (GGM) filtered image of the merger, as shown in the figure \ref{fig:ggm} in the $0.5-7.0$ keV energy range. GGM filtering is a robust edge-detection technique which is very useful in resolving the substructures in a cluster core, as well as at cluster outskirts. This filter calculates the gradient of an image assuming Gaussian derivatives, with the intensity of the GGM images indicating the slope of the local surface brightness gradient, where steeper gradients show up as brighter regions \citep{Sanders_2016_ggm}. GGM images have been utilized in various scientific fields, including physics and astronomy, to map substructures with great visible clarity \citep{Walker_2016_ggm}.

For the presented GGM image in Fig. \ref{fig:ggm}, we applied a 3-pixel scale filter, binning the \textit{Chandra} image by a factor of 2 to yield pixels of width $0.949"$. This GGM image reveals two surfaces with pronounced brightness gradients: the Primary shock and SE edge, delineated by dashed red lines in Figure \ref{fig:ggm}. In order to further investigate these edge features, we used spatial spectroscopy techniques. 

\section{Spatially Resolved Spectroscopy} \label{spatial}

\begin{figure*}	
	\includegraphics[width=\linewidth]{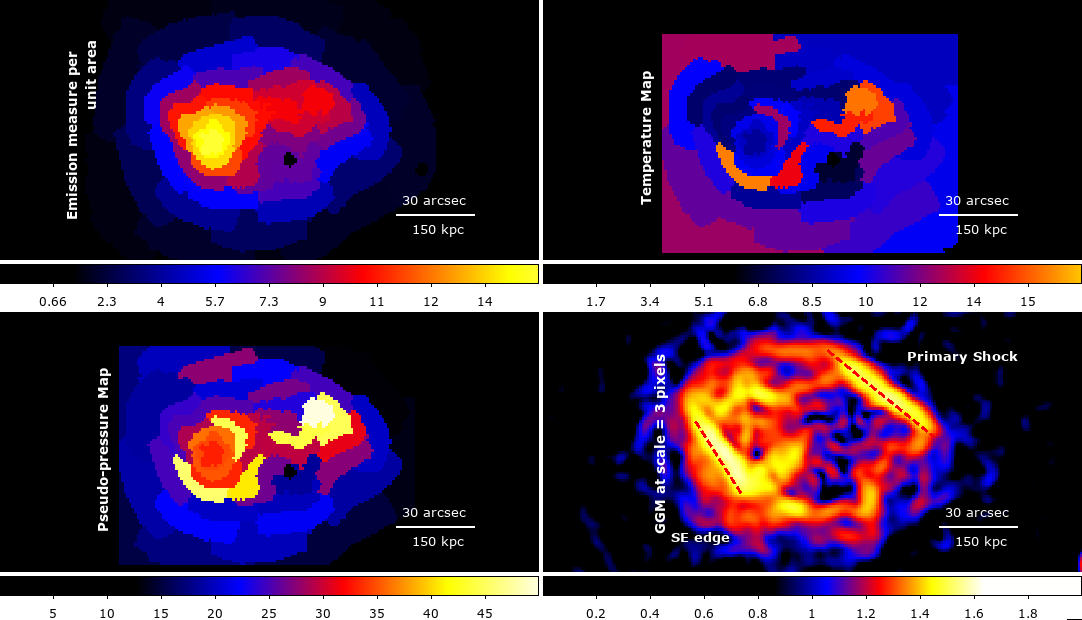}
    \caption{Top left: Projected emission per unit area (cm$^{-5}$ arcsec$^{-2}$. Top Right: Projected Temperature Map (keV) with S/N = 32. Bottom Left: Projected pseudo-pressure map (keV cm$^{-5}$ arcsec$^{-2}$), obtained by multiplying the emission measure and temperature maps. The small black circles in the emission and temperature map, and the pseudo pressure map are the excluded point sources. Bottom Right: GGM image of SPT J2031 in the $0.5-7.0$ keV energy range at scale = 3 pixels. }
    \label{em-meas_tmap_press_ggm-5}
\end{figure*}

Spatially resolved spectroscopy techniques were used to produce maps of projected gas properties of the cluster (see Fig. \ref{em-meas_tmap_press_ggm-5}). The central $\sim 3 \times 3$ arcmin region was divided into bins using the Contour Binning algorithm \citep{Sanders_contour_binning} which creates bins based on the variations in surface brightness. The signal-to-noise ratio was chosen to be 32 ($\sim$ 1000 counts) for obtaining the bins, as was used in \cite{Russell_2012}. For all the 66 regions obtained this way, spectra were extracted for each observation and appropriate RMFs and ARFs were generated. 
The background for each of these spectra was subtracted using the normalized blank-sky backgrounds, as discussed in section \ref{data_reduction}.
These spectra were restricted to the energy range of 0.5 - 7.0 keV. The spectra for each region were then simultaneously fitted for all observations using Sherpa with the PHABS(APEC) model, where the hydrogen column density is fixed at $n_H = 3.0 \times 10^{20}$ cm$^{-2}$ \citep{NH_Kalberla}, the solar abundance is $0.3$ Z$_\odot$, and the redshift is $0.34$ and C statistics were applied. The pseudo-pressure map is produced by multiplying the square root of the emission measure and temperature maps.

The panels in the top row and the bottom left of figure \ref{em-meas_tmap_press_ggm-5} show the projected emission per unit area map (cm$^{-5}$ arcsec$^{-2}$), projected temperature map (keV), projected pseudo pressure map (keV cm$^{-5}$ arcsec$^{-2}$)  from left to right, while the bottom right panel is the GGM image at a scale of 3 pixels.


Each edge in the GGM image corresponds to a jump in temperature in the temperature map and a jump in pressure in the pseudo-pressure map. This makes the edges consistent with being shock fronts.

\section{Shock Fronts} \label{shock}

While a number of clusters have been found to have shock-heated regions, the detection of a cluster merger with sharp surface brightness edges and a distinctive high-temperature jump is rare due to the requirement of favourable merger geometry \citep{Zuhone_2022}. In fact, only a handful of merger shock fronts with a high Mach number,  $ M > 2.0$ have been discovered by \textit{Chandra}, such as the Bullet Cluster with $ M = 3.0 \pm 0.4 $ \citep{Markevitch_2006_BC}, A2146 with $ M = 2.3 \pm 0.2  $ \citep{Russell_2010,Russell_2012,Russell_2022}, A665 with $ M = 3.0 \pm 0.6  $ \citep{Dasadia_2016}, El Gordo with $ M \geq 3 $ \citep{Botteon_2016_El_gordo}, A520 with $ M = 2.4_{-0.3}^{+0.4}  $ \citep{wang_2018}, and A98 with $M = 2.3 \pm 0.3 $ \citep{Sarkar_2022}. \textsl{Chandra} has also determined shock fronts with $ M < 2.0$, such as A2744 with $ M = 1.41_{-0.08} ^{+0.13}$  \citep{Owers_2011_A2744}, A754 with $ M = 1.57 _{-0.12}^{+0.16}$ \citep{Macario_2011_A754}, A521 with $ M = 2.4 \pm 0.2 $ \citep{Bourdin_2013} and A2034 with $ M = 1.59_{-0.07}^{+0.06}$ \citep{Owers_2014_A2034}. 

To determine the Mach number for SPT J2031, we extracted surface brightness profiles and temperature profiles across both the shock fronts.  
\begin{figure*}
	\includegraphics[width=1.0
	\linewidth]{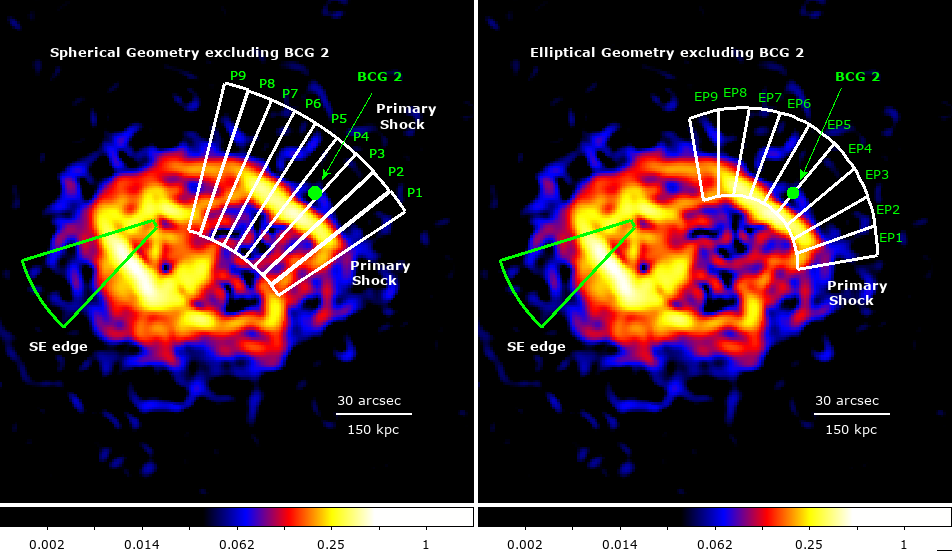}
	
    \caption{GGM image at scale=3 pixels showing the sectors used to extract surface brightness profiles assuming spherical geometry (\textit{left}) and elliptical geometry (\textit{right}). The sectors in white are used to measure the density jump across the primary shock and the one in green for the SE edge. The filled green circle is the BCG2 which is excluded while extracting the surface brightness profiles. }
    
    \label{fig:ggm_3-all_sectors}
\end{figure*}

\subsection{Surface Brightness Profiles} \label{sb_profiles}
Sharp discontinuities in the X-ray surface brightness are observed in Fig. \ref{fig:spt_no_pt_srcs_labelled} and in the top left panel of Fig. \ref{em-meas_tmap_press_ggm-5}. In order to investigate these discontinuities, we extracted surface brightness profiles in the northwestern region covering the primary shock front and in the southeastern region covering the SE edge. Consistent with previous studies (Russell, 2010; Russell, 2012; Russell, 2022), our initial analysis assumed spherical geometry to derive these profiles. However, in subsequent analysis, we also explored an alternative approach by extracting the profiles using elliptical annuli as seen in \cite{Ogrean_2014}. This additional analysis aimed to ascertain whether an elliptical geometry provides a more accurate description of the shock geometry.

	
    

\subsubsection{Spherical Geometry}

\begin{figure*}
	\hbox{
	\includegraphics[width=0.33\linewidth]{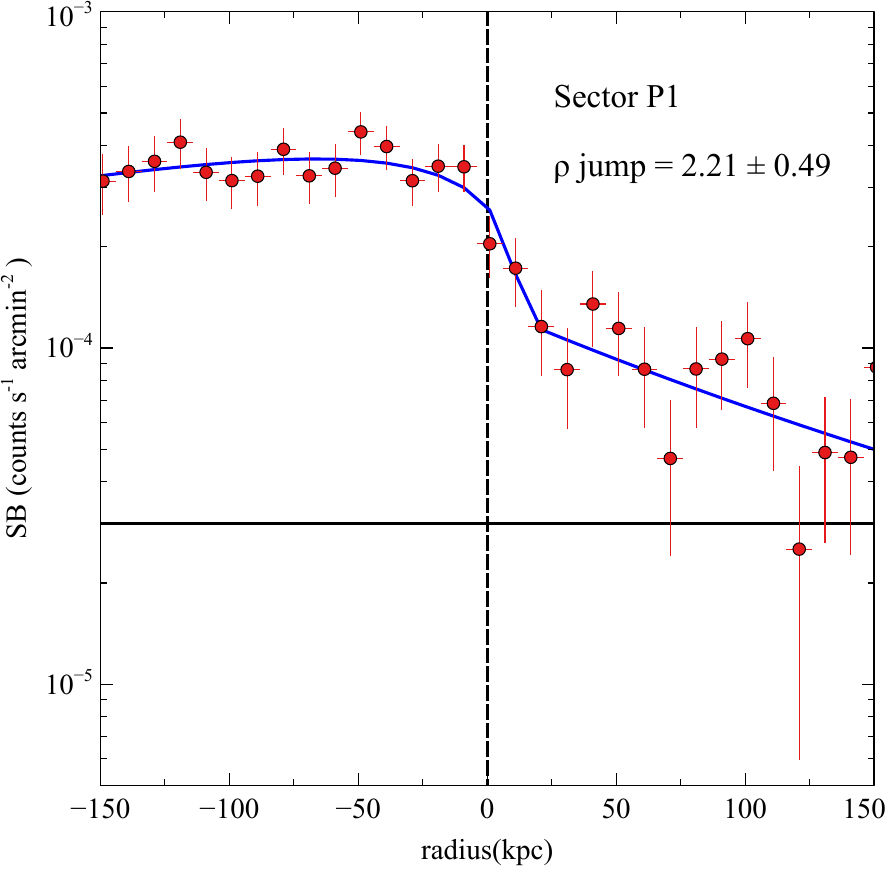}
	\includegraphics[width=0.33\linewidth]{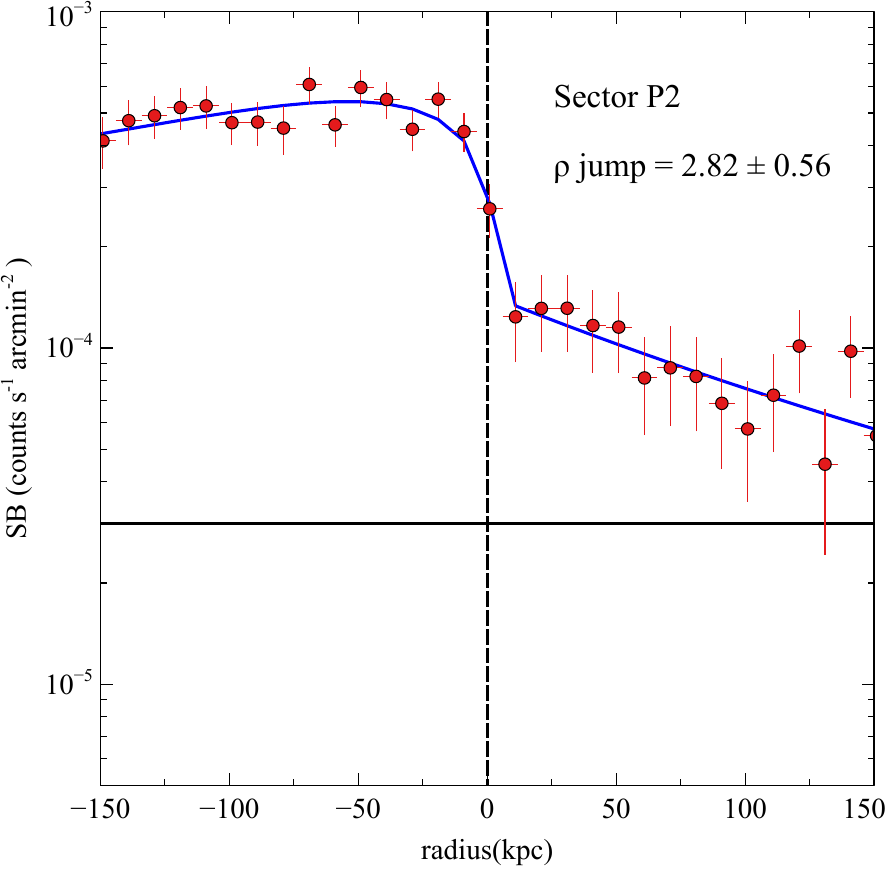}
    \includegraphics[width=0.33\linewidth]{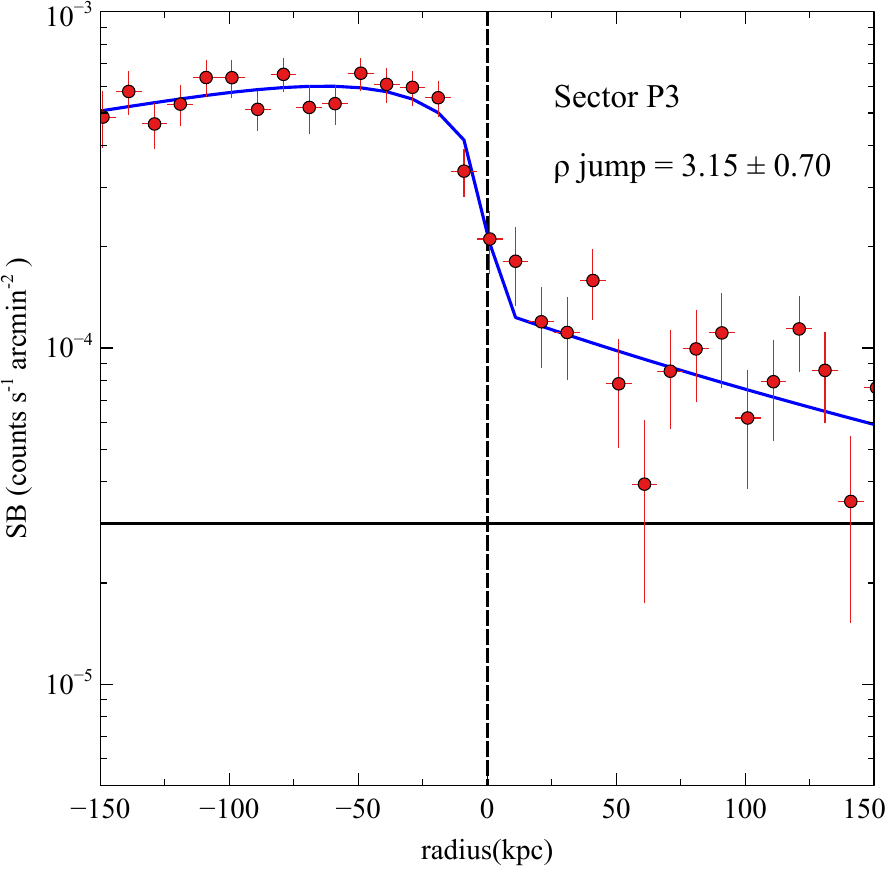}
	}
    \hbox{
	\includegraphics[width=0.33\linewidth]{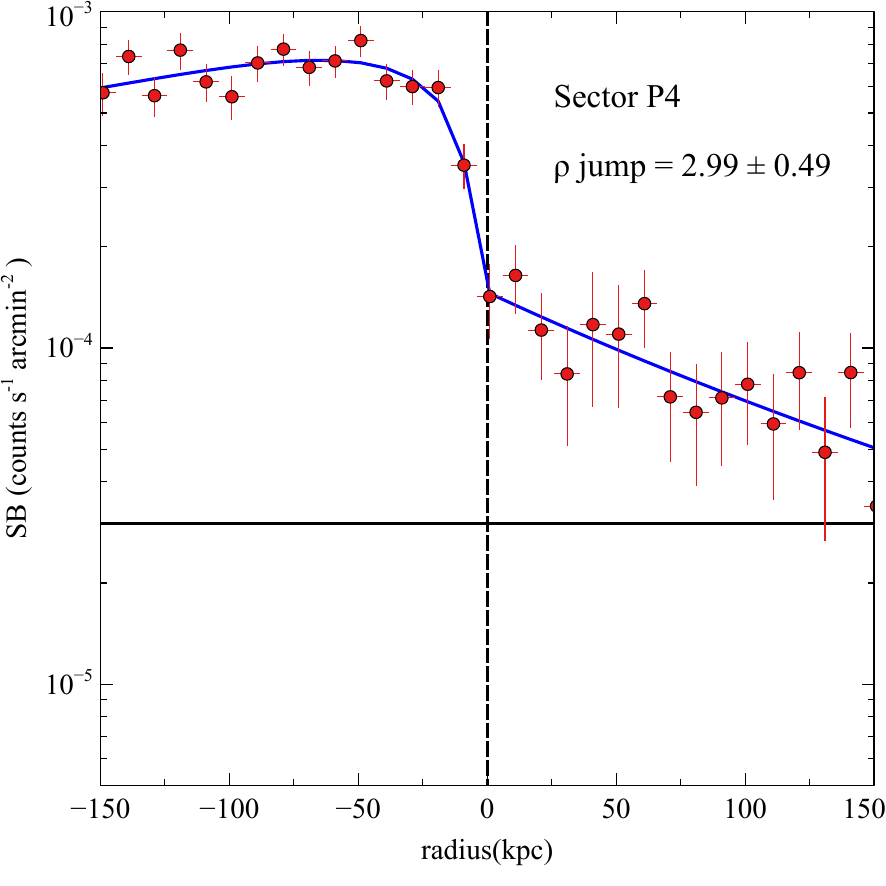}
	\includegraphics[width=0.33\linewidth]{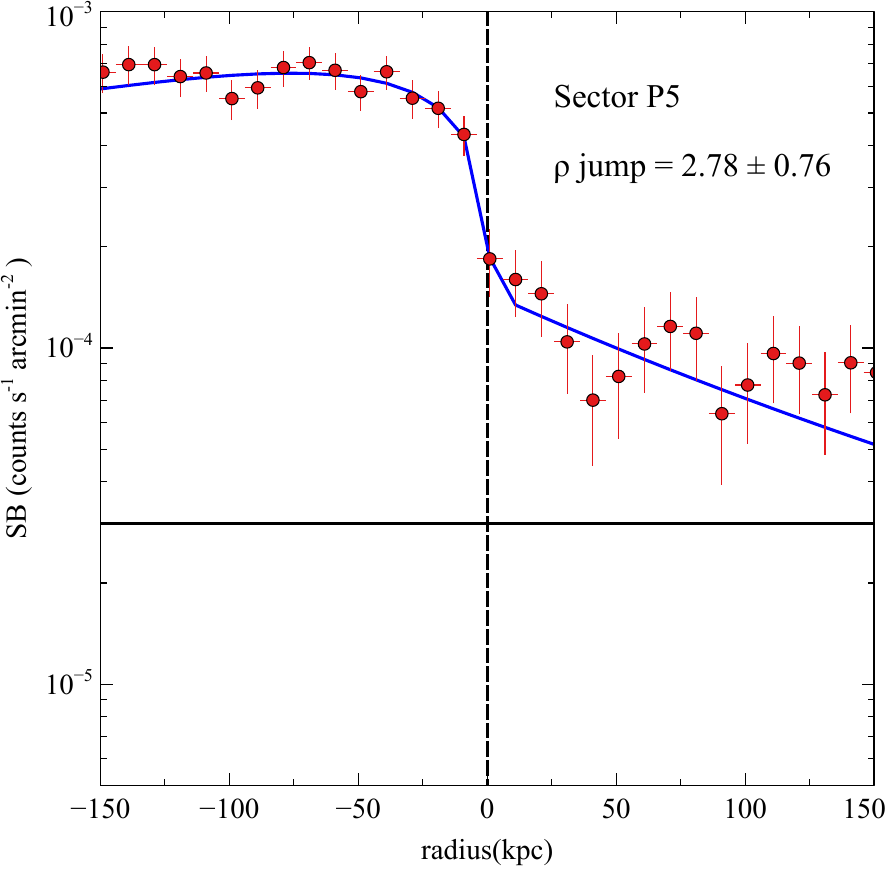}
    \includegraphics[width=0.33\linewidth]{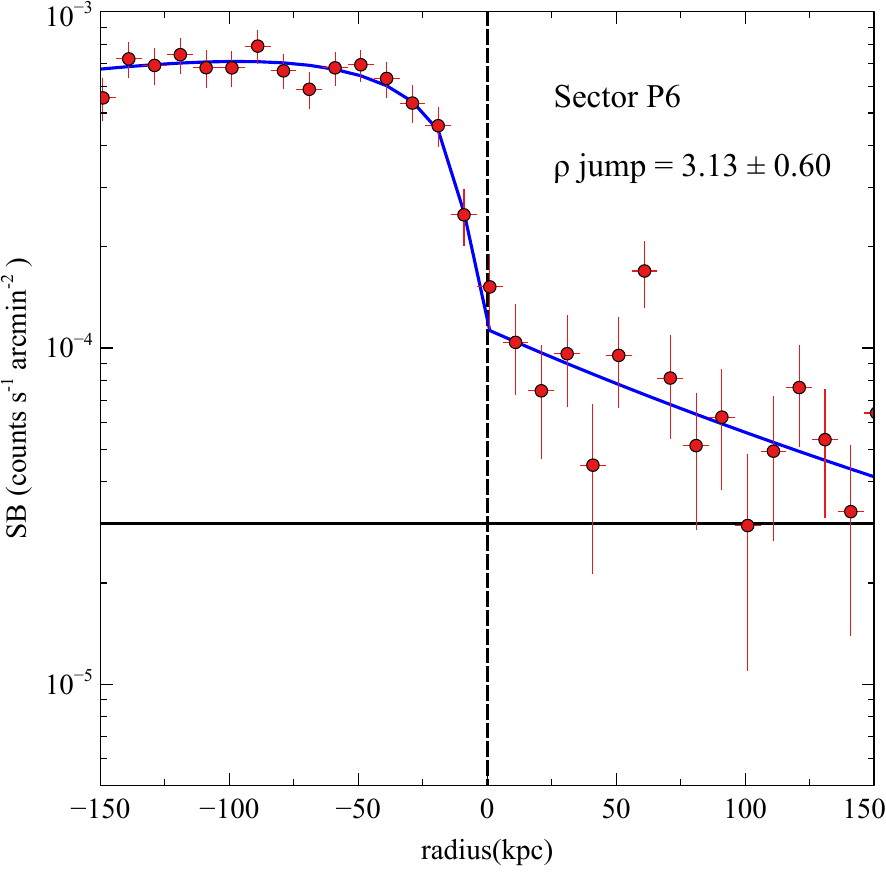}
	}
    \hbox{
	\includegraphics[width=0.33\linewidth]{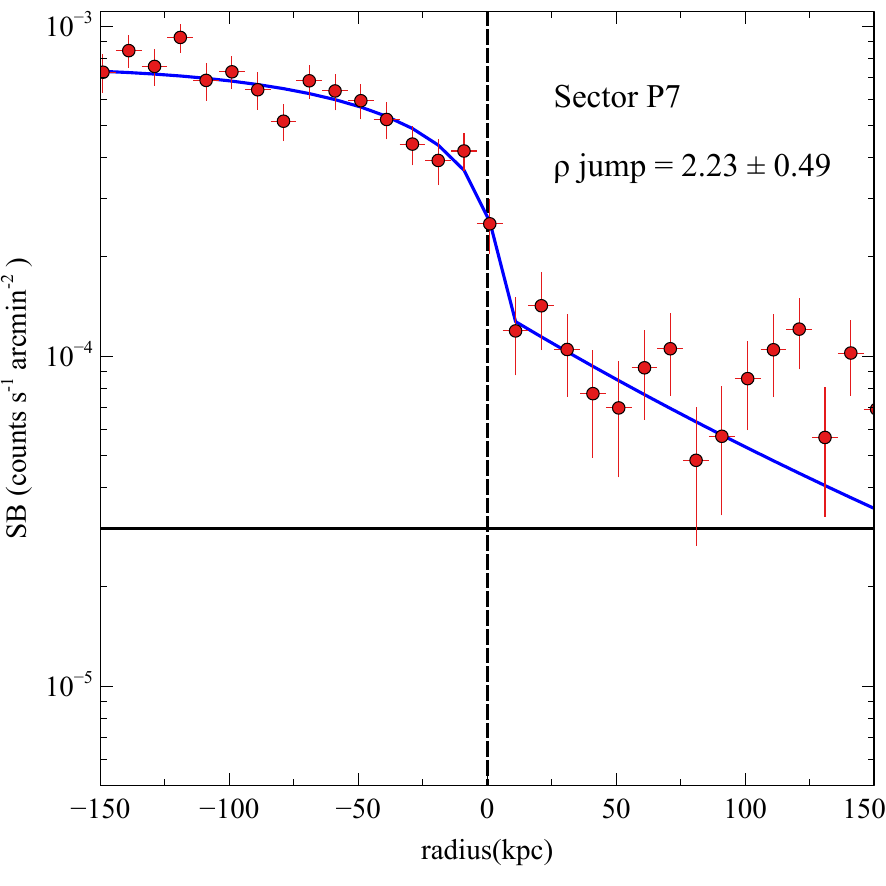}
	\includegraphics[width=0.33\linewidth]{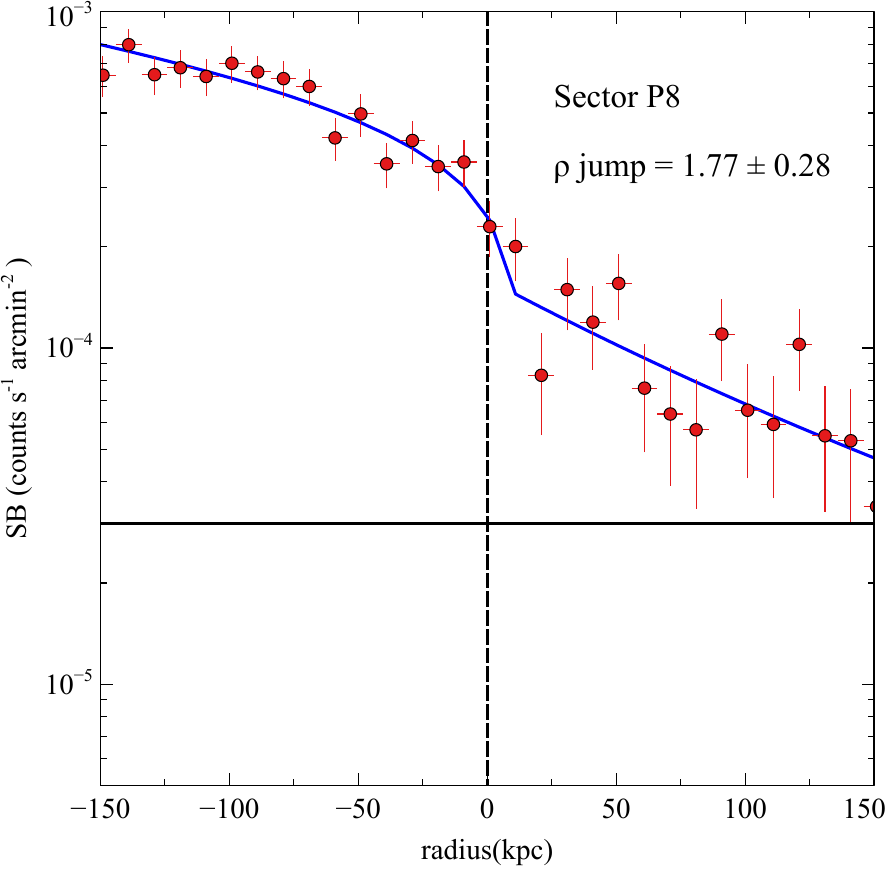}
    \includegraphics[width=0.33\linewidth]{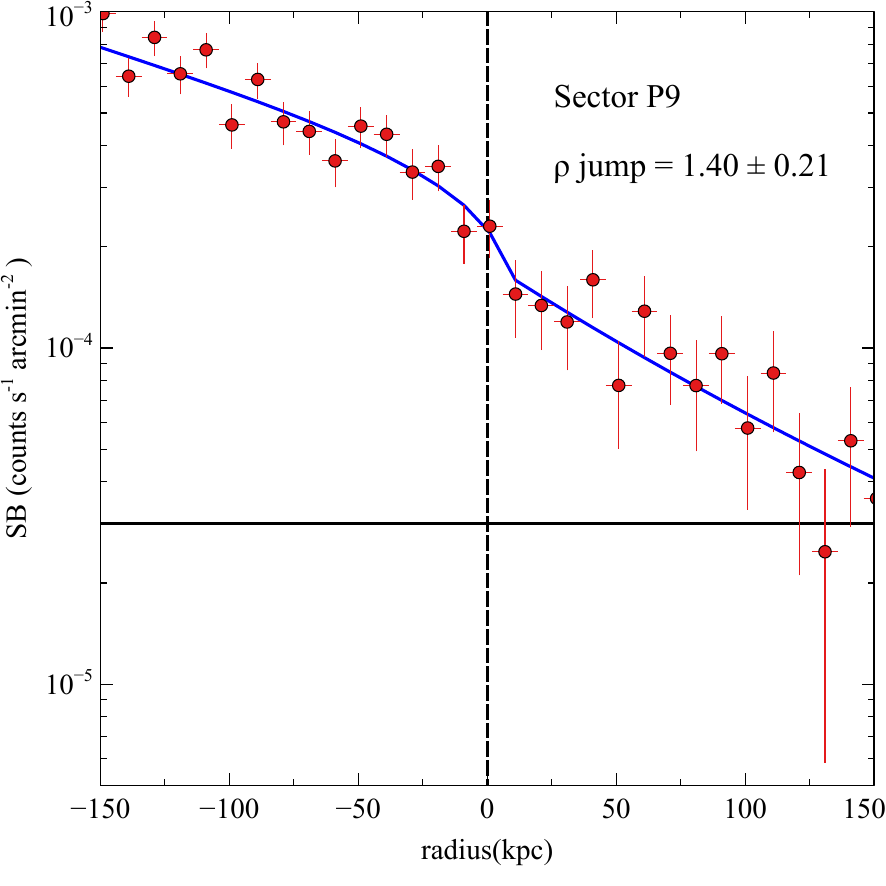}
	}
    \caption{Surface brightness profiles in the $0.5 - 2.5$keV energy band across sectors P1-9, each background subtracted (solid black) and fitted with the broken power law density model (in blue). }
    \label{fig:ps-1-9-sb-spherical}
\end{figure*}

\begin{figure*}
	\hbox{
	\includegraphics[width=0.33\linewidth]{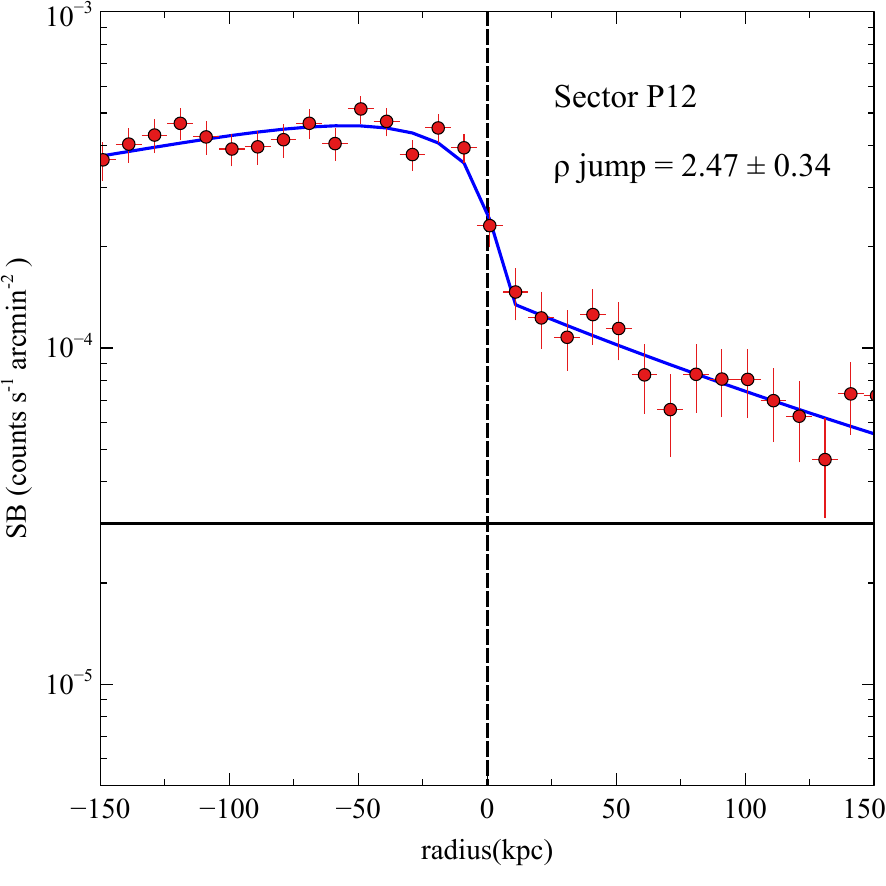}
	\includegraphics[width=0.33\linewidth]{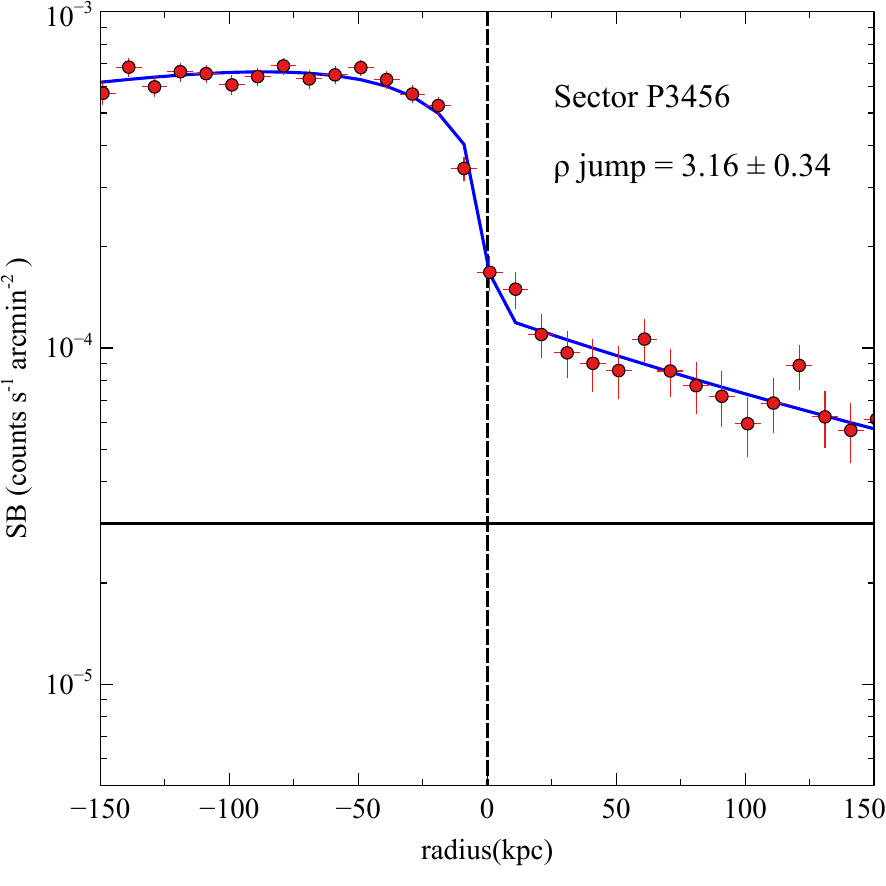}
    \includegraphics[width=0.33\linewidth]{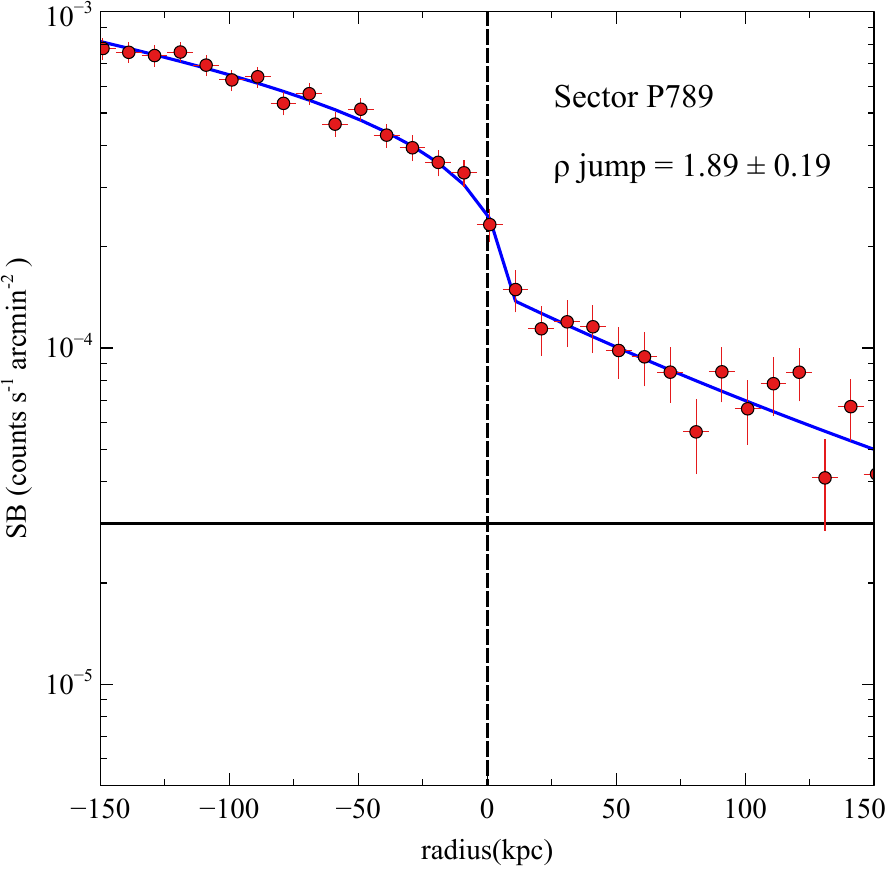}
	}
    
    \caption{Surface brightness profiles across sectors P1-2 (left panel), P3-6 (center panel) and P7-9 (right panel) in the $0.5 - 2.5$ keV energy band. Each profile has been background-subtracted and fitted with the broken power law density model (in blue). }
    \label{fig:ps-new-all-sb}
\end{figure*}

 The left panel of Fig. \ref{fig:ggm_3-all_sectors} shows the sectors selected for extracting surface brightness profiles assuming a spherical geometry. The sectors are chosen to cover the region where the shock fronts are well-defined, based on the GGM image. Sectors $P1 - P9$ (where P refers to the primary shock) extend over the primary shock front. These sectors are centered to fully analyse the jump in surface brightness. The outer radius for each sector was taken to be 5 arcmin (although the sectors in the image extend only up to $\sim$ 1.5 arcmin). Taking into account that BCG2 lies in the direction of these sectors, we excluded the region containing BCG2 before extracting the surface brightness profiles. This was done to prevent any potential impact from BCG2 on our analysis.

Once extracted, the surface brightness profiles were fitted with a broken power-law model projected along the line of sight \citep{MV_review_2007} with the aim of identifying density discontinuities in the chosen sectors. Assuming spherical symmetry (following \citealt{Russell_2012}), the density distribution can be given by:

\begin{equation}
    n(r)=
    \begin{cases}
       n_0 \Bigl(\frac{r}{r_{sh}}\Bigr)^{\alpha_1},  &\text{if r $\leq $ r$_{sh}$} \\
       \frac{1}{C}n_0 \Bigl(\frac{r}{r_{sh}}\Bigr)^{\alpha_2},  &\text{if r $>$  r$_{sh}$} \\
    \end{cases}
\end{equation}

\noindent where $n_0$ is the density normalisation, $\alpha_1$ and $\alpha_2$ are the power-law indices, $r_{sh}$ is the assumed shock location where the discontinuity in the surface brightness occurs. Also $C = \rho_2/\rho_1$, where $\rho_2$ is the post-shock density and $\rho_1$ is the pre-shock density. At the location of the shock, $\rho_2$ is greater than $\rho_1$, \citep{mirakhor_2023}. 

Using Rankine-Hugoniot jump conditions for the density jump, the Mach number for each sector can be calculated as follows: 

\begin{equation}
    M = \left[\frac{2 \frac{\rho_2} { \rho_1}}{\gamma + 1 - (\frac{\rho_2}{\rho_1})(\gamma -1)}\right]^\frac{1}{2}
    \label{M_dens_jump}
 \end{equation}
 
 \noindent where $\rho_{2}/\rho_1$ is the density jump, and $\gamma = 5/3$ for a monoatomic gas \citep{Russell_2010}.

In Fig. \ref{fig:ps-1-9-sb-spherical}, the red crosses in each panel show the surface brightness profile across sectors P1-P9. For each of these sectors, there is a sharp discontinuity in the surface brightness. The regions to the right of this jump are the pre-shock regions and the ones to the left, with the higher surface brightness are the post-shock regions. 
	

\begin{table*}
\centering
\caption{\label{density_jump_sector} Details of the surface brightness fitting across the sectors along the primary shock front. The columns are, from left to right: sector label, density jump across that sector obtained by fitting with the broken power law density model, Mach number obtained from the density jump, the inner and outer slopes (power law indices in the broken power law model) and the reduced chi-squared of the fit.}
\begin{tabular}{|l|l|l|l|l|l|}

\hline
Sector  & Density Jump & M & $\alpha_1$ & $\alpha_2$ & $\chi^{2}/\nu$  \\ [0.5ex] 

\hline
$P$$1$ & $ 2.21\pm 0.55$ & $1.92_{-0.29}^{+0.43}$ & $-1.80\pm 0.83 $ & $2.17\pm 0.63$ & 19.62/25  \\
\hline
$P$$2$ & $2.8 \pm 0.52 $ & $2.65_{-0.83}^{+1.39}$ & $-2.67 \pm 0.65$ & $2.08 \pm 0.50$ & 14.11/25 \\
\hline
$P$$3$ & $3.15 \pm 0.7$ & $3.35_{-0.85}^{+1.95}$& $-2.33 \pm 0.48 $ & $1.87 \pm 0.55$  & $27.49/25$ \\
\hline
$P$$4$ & $2.99 \pm 0.49$ & $2.99_{-0.52}^{+1.03}$ & $ -2.55 \pm 0.55$ & $2.33 \pm 0.47$ & 17.65/25 \\
\hline
$P$$5$ & $2.78 \pm 0.77$ & $2.63_{-0.62}^{+1.41}$ & $-1.89 \pm 0.29$ & $2.30 \pm 1.01 $ & $13.98/25$ \\
\hline
$P$$6$ & $3.13 \pm 0.6$ & $3.3_{-0.73}^{+1.91}$ & $ -1.55 \pm 0.44$ & $ 2.23 \pm 0.55 $ & $20.43/25$ \\
\hline
$P$$7$ & $2.23 \pm 0.49$ & $1.94_{-0.28}^{+0.43}$ & $-4.85 \pm 0.2$ & $2.94 \pm 1.01 $ & $23.14/25$ \\
\hline
$P$$8$ & $1.77 \pm 0.29$ & $1.54_{-0.14}^{+0.18}$ & $0.53 \pm 0.38$ & $2.60 \pm 0.53 $ & $23.32/25$ \\
\hline
$P$$9$ & $1.4 \pm 0.21$ & $1.27_{-0.09}^{+0.12}$ & $0.95 \pm 0.39$ & $3.04 \pm 0.52 $ & $23.28/25$ \\
\hline
$P$$1-2$ & $2.47 \pm 0.34$ & $2.2_{-0.23}^{+0.35}$ & $-2.56 \pm 0.51$ & $2.16 \pm 0.37 $ & $17.37/25$ \\
\hline
$P$$3-6$ & $3.16 \pm 0.34$ & $3.36_{-0.48}^{+0.87}$ & $-1.63 \pm 0.19$ & $1.86 \pm 0.27 $ & $23.28/25$ \\
\hline
$P$$7-9$ & $1.89 \pm 0.19$ & $1.64_{-0.09}^{+0.12}$ & $0.53 \pm 0.23$ & $2.40 \pm 0.30 $ & $16.86/25$\\
\hline
SE edge & $1.53 \pm 0.14$ & $1.36_{-0.08}^{+0.09}$ & $-1.17 \pm 0.43$ & $1.45 \pm 0.39 $ & $67.13/64$\\
\hline
\end{tabular}

\end{table*}

We have binned the sectors P1-2, P3-6, and P7-9 so that the sectors containing the part of the edge with the highest density jump (P3-6) are binned together. This binning allows us to better constrain the values of the density jumps, and how the density jump varies along the shock front. Sector P3-6 is designed to cover the steepest part of the jump based on the GGM image, while the regions P1-2 and P7-9 cover the regions to either side of the steepest jump.  All three plots in Fig. \ref{fig:ps-new-all-sb} show the surface brightness profiles over sectors P1-2, P3-6, and P7-9 fitted with the broken power law model in indicated by the solid blue line. Table \ref{density_jump_sector} shows the power law indices and density jump obtained from this fitting. The table shows these values for all the individual sectors P1-9, the binned sectors P1-2, P3-6, P7-9 and the SE edge.


\begin{figure*}
	\hbox{
	\includegraphics[width=0.5\linewidth]{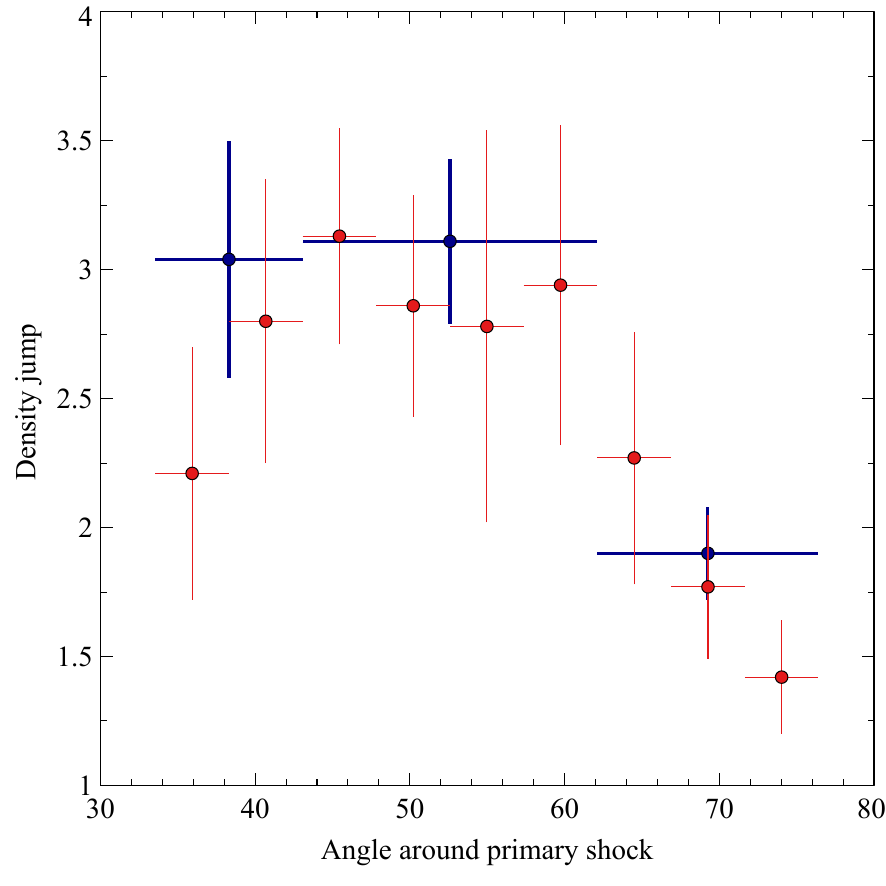}
	\includegraphics[width=0.5\linewidth]{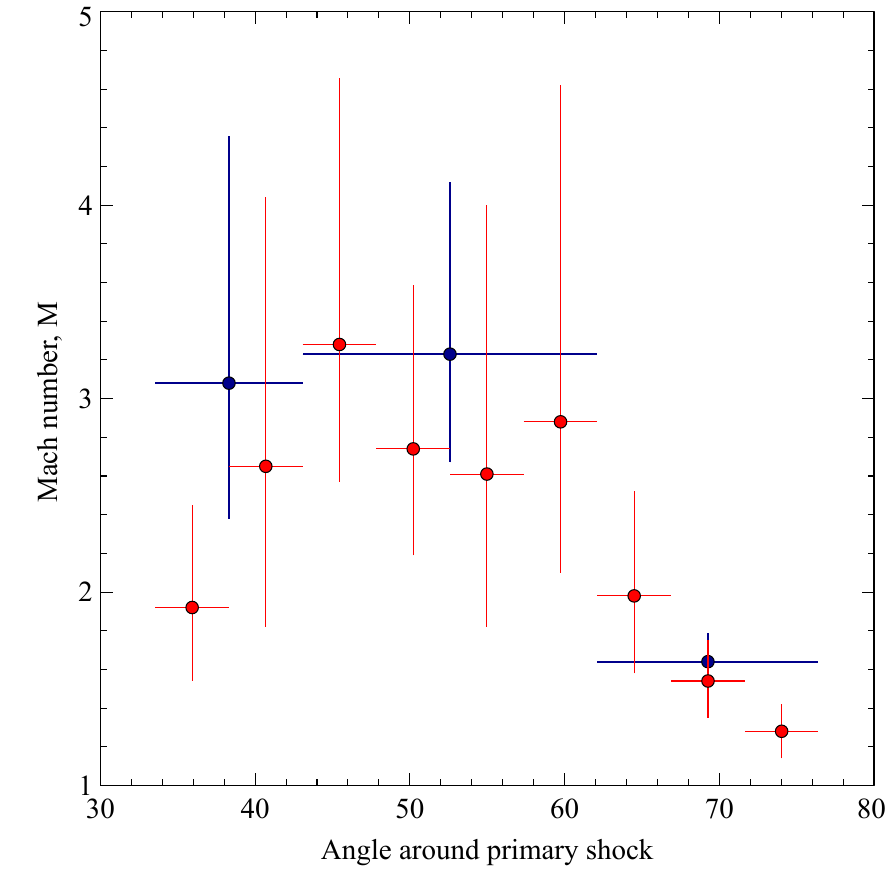}
	}
    \caption{ \textit{Left:} The cross-bars in red depict the density jump (spherical geometry) across each of the sectors P1 - P9 and the ones in blue represent the density jumps in sectors P1-2, P3-6 and P7-9 from left to right. The x-axis represents the angle of the sectors around the primary shock front, going from $33^{\circ}$ to $77^{\circ}$. \textit{Right:} The Mach number, determined from the density jump, is shown here in red for the sectors P1-9 and for sectors P1-2, P3-6, and P7-9 shown in blue, across the angles of the sectors around the primary shock front.    }
    \label{fig:dj_m_vs_angles}
\end{figure*}

\
The density jump and Mach number obtained from sectors P1-9 along the Primary shock are plotted in Fig. \ref{fig:dj_m_vs_angles}.
 The panel on the left shows the density jumps across the sectors plotted against the angle around the primary shock front going from $33^{\circ}$ to $77^{\circ}$. The red crosses indicate the density jump across each individual sector P1-9. The blue crosses represent the sectors binned as P1-2, P3-6, and P7-9. The panel on the right shows the Mach numbers derived from the corresponding density jumps using equation \ref{M_dens_jump}, also plotted against the angle around the primary shock front. The red crosses represent the Mach number derived for each of the sectors P1-9 and the blue crosses represent the binned sectors P1-2, P3-6, and P7-9. 

In both the plots, the observed trend is that the density jump and Mach number are highest at the center of the shock front where the GGM image shows the highest gradient.  On both sides of this center point, as the brightness of the GGM image decreases, the values of density jump and Mach number taper off, as expected from a similar analysis performed in \cite{Russell_2022}. The peak values of the density jump and Mach number are $3.11 \pm 0.32$ and $3.23_{-0.56}^{+0.89}$ respectively for the binned sector P3-6, the brightest region in the GGM image.

\subsubsection{Elliptical Geometry}

\begin{figure*}
	\hbox{
	\includegraphics[width=0.33\linewidth]{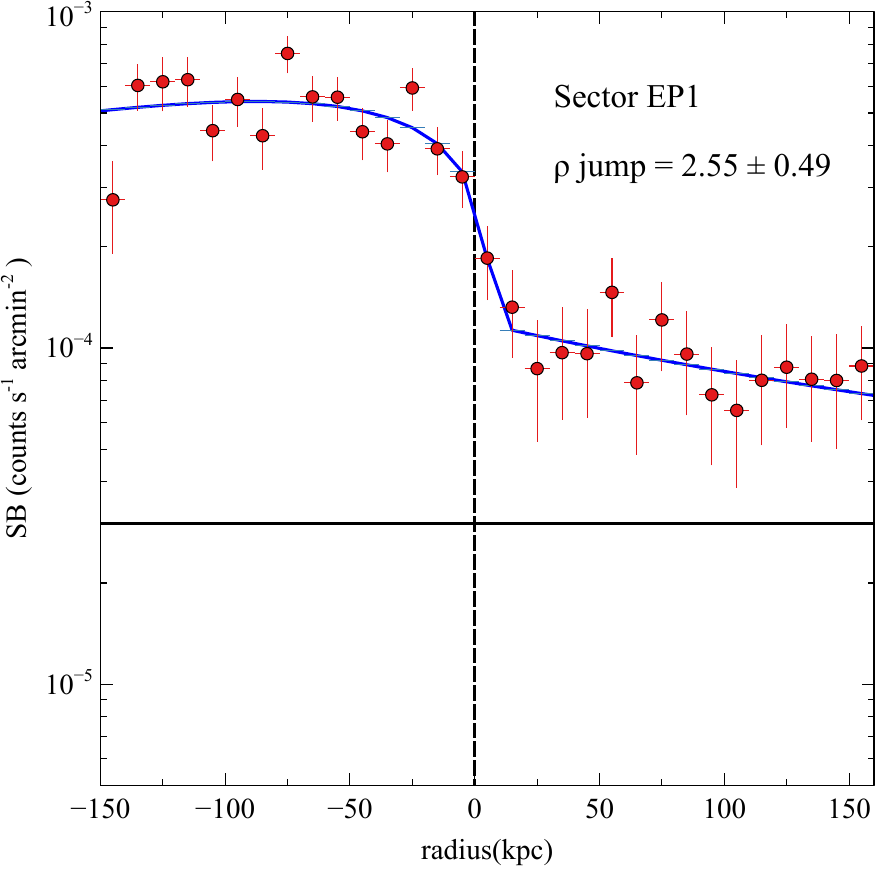}
	\includegraphics[width=0.33\linewidth]{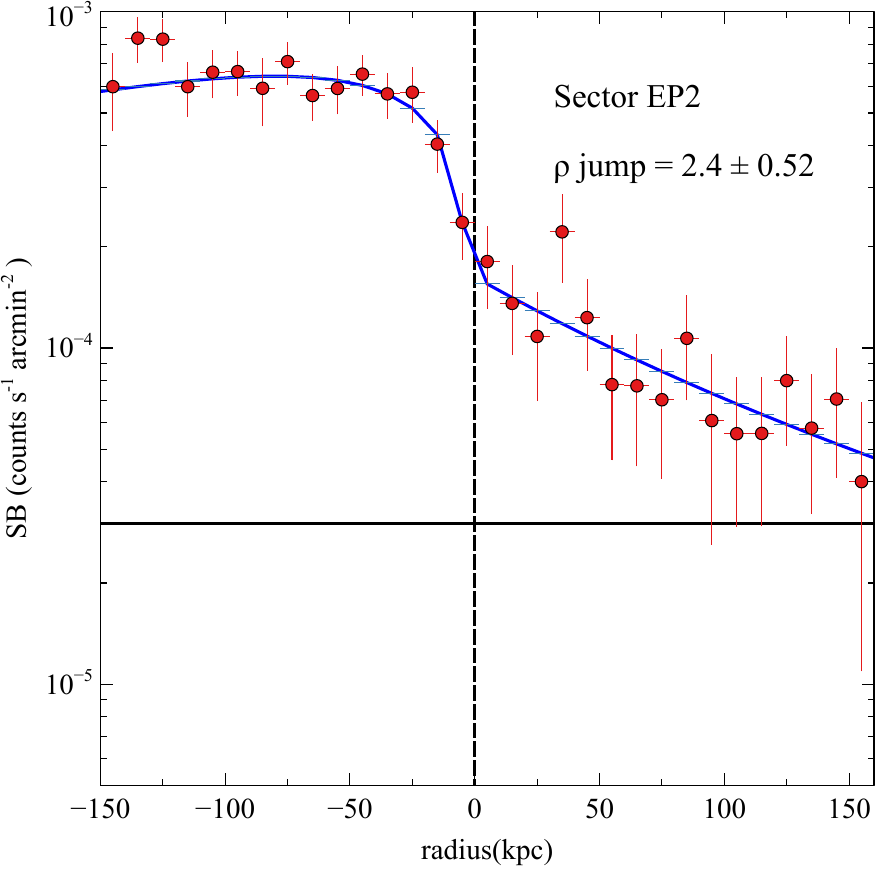}
    \includegraphics[width=0.33\linewidth]{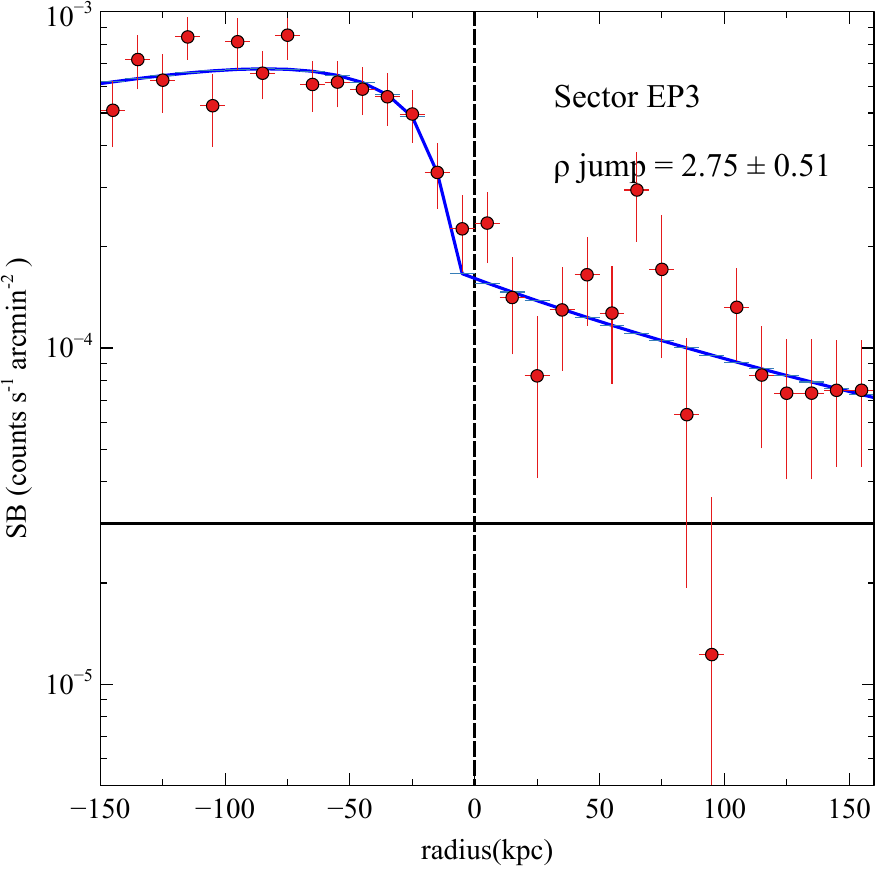}
	}
    \hbox{
	\includegraphics[width=0.33\linewidth]{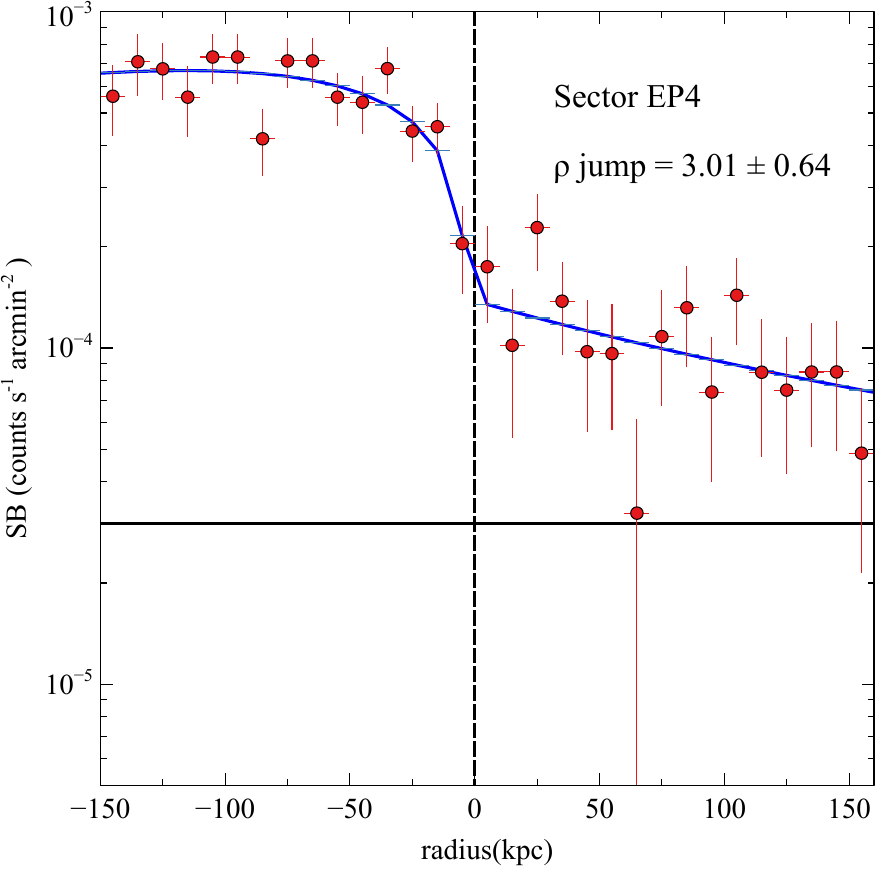}
	\includegraphics[width=0.33\linewidth]{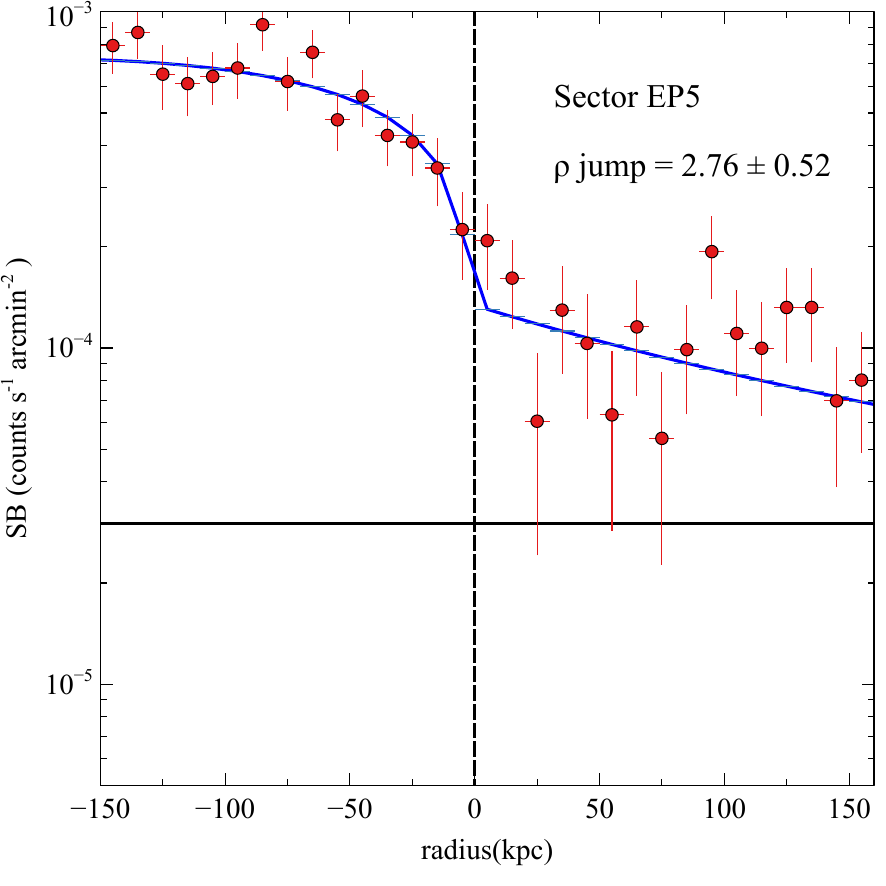}
    \includegraphics[width=0.33\linewidth]{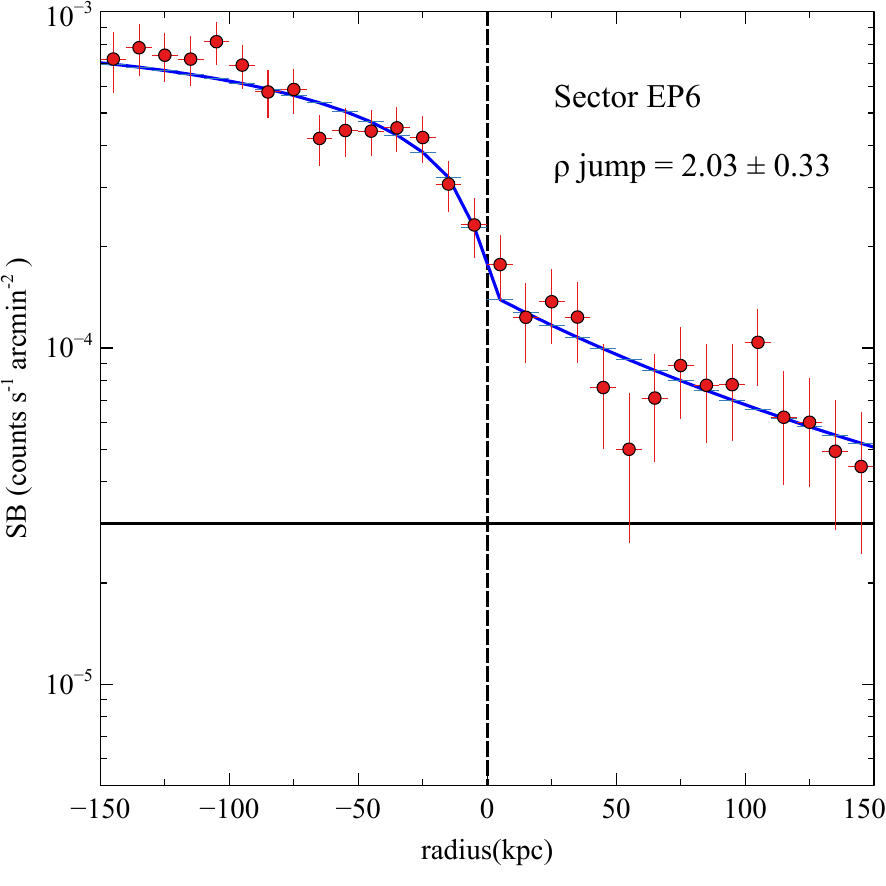}
	}
    \hbox{
	\includegraphics[width=0.33\linewidth]{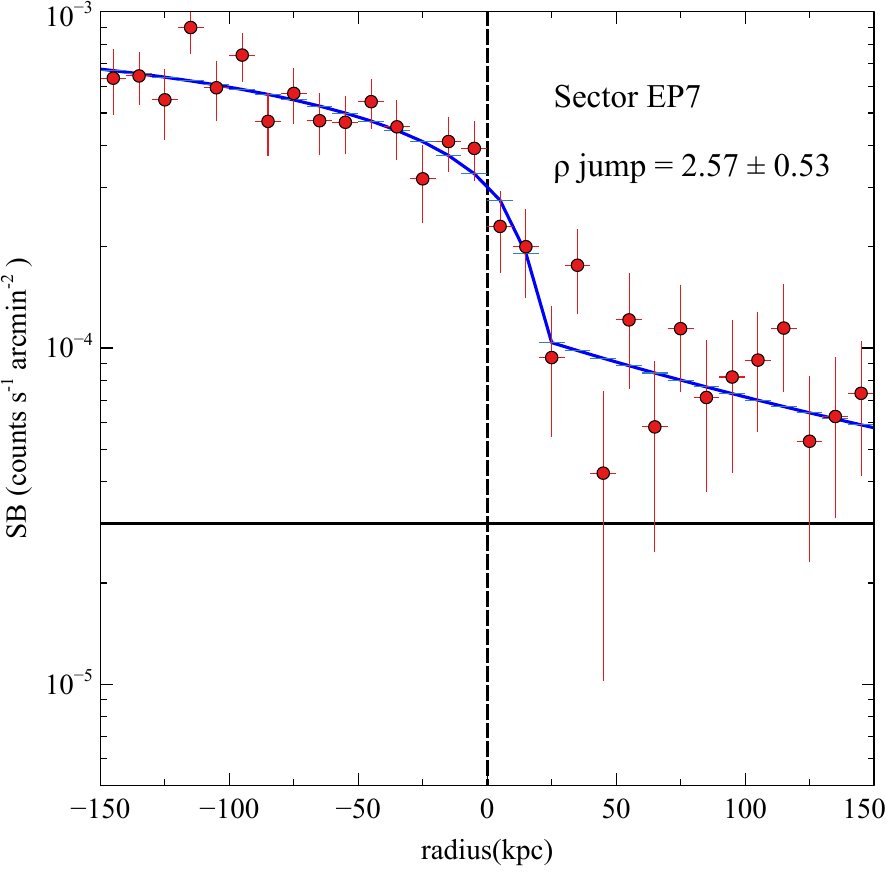}
	\includegraphics[width=0.33\linewidth]{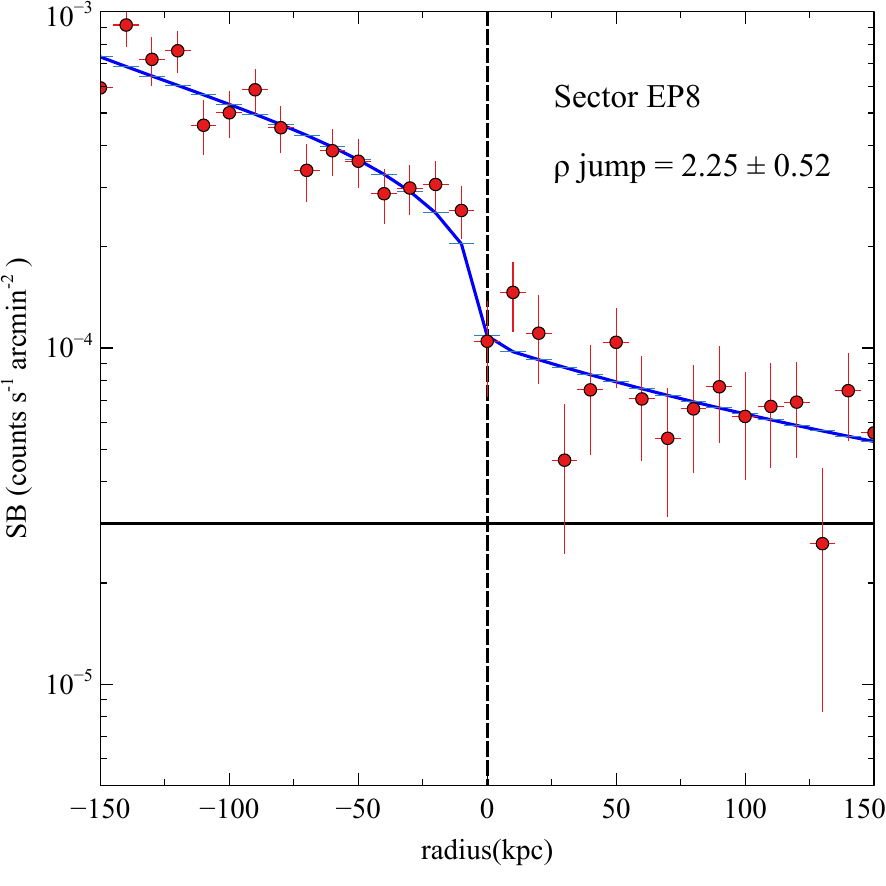}
    \includegraphics[width=0.33\linewidth]{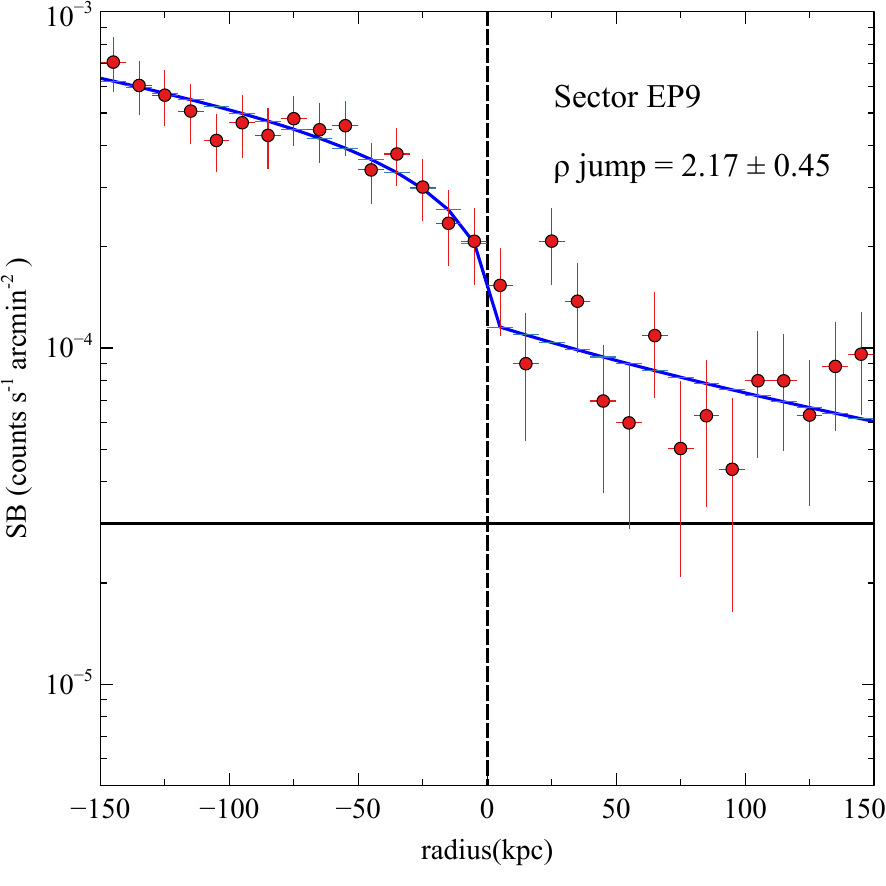}
	}
    \caption{Surface brightness profiles (in red) in the $0.5-2.5$ keV energy band across individual elliptical sectors EP1-9, fitted with the broken power law density model (in blue) after performing background subtraction (solid black). }
    \label{fig:ep-1-9-sb}
\end{figure*}

\begin{figure*}
	\hbox{
	\includegraphics[width=0.33\linewidth]{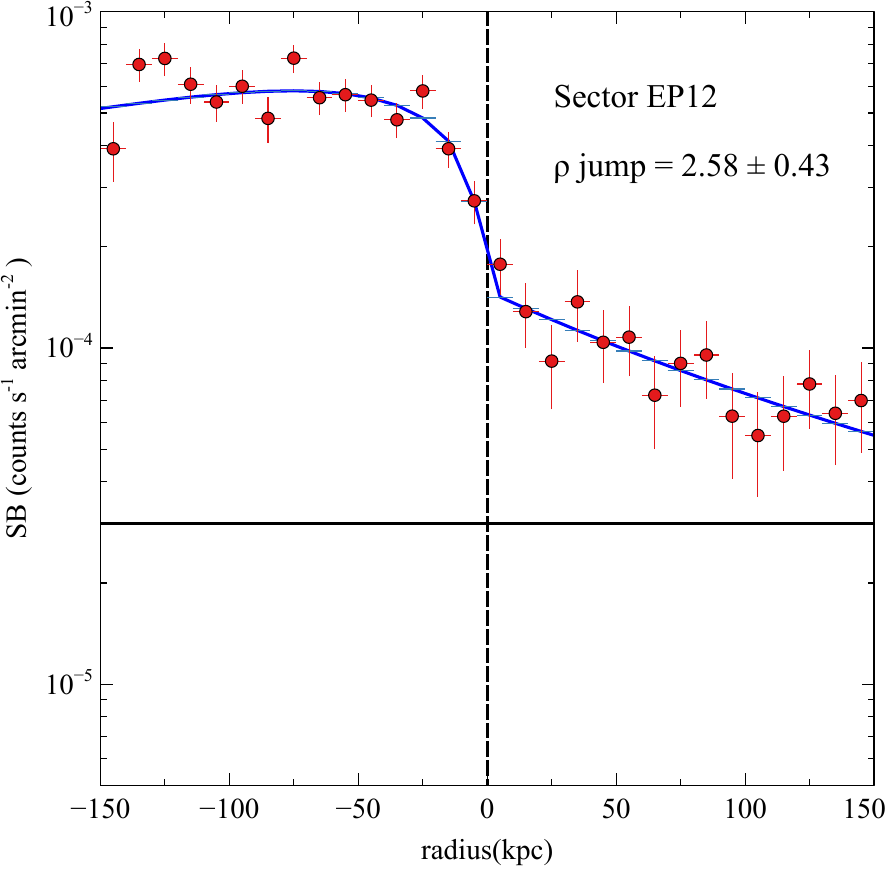}
	\includegraphics[width=0.33\linewidth]{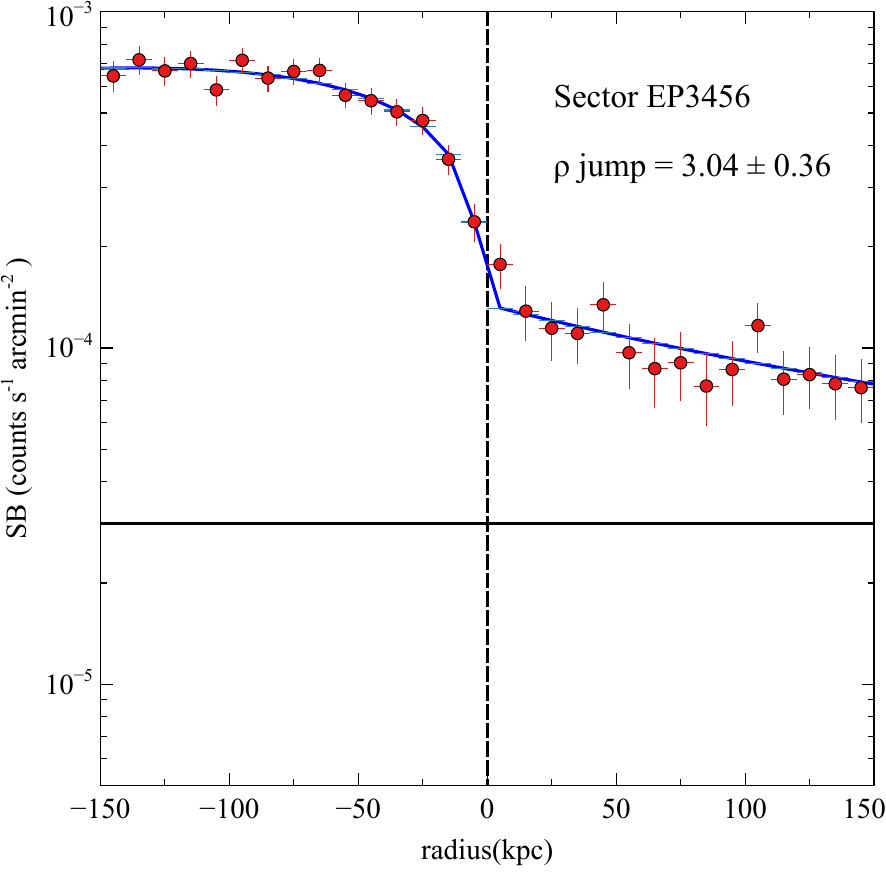}
    \includegraphics[width=0.33\linewidth]{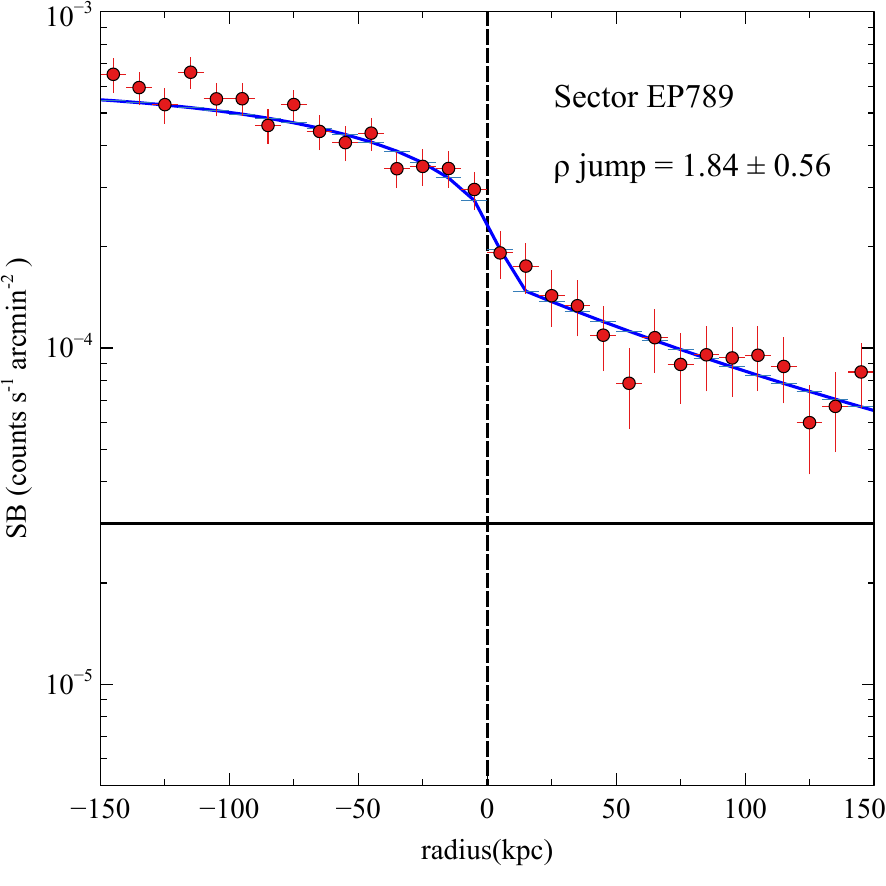}
	}
    
    \caption{Surface brightness profiles across sectors with elliptical geometry EP1-2 (left panel), EP3-6 (center panel) and EP7-9 (right panel) in the $0.5 - 2.5$ keV energy band. Each profile has been background-subtracted and fitted with the broken power law density model (in blue). }
    \label{fig:ep_binned_-sb}
\end{figure*}

\begin{table*}[h]
\centering
\caption{\label{density_jump_sector_elliptical} Details of the surface brightness fitting across the sectors along the primary shock front. The columns are, from left to right: sector label, density jump across that sector obtained by fitting with the broken power law density model, Mach number obtained from the density jump, the inner and outer slopes (power law indices in the broken power law model) and the reduced chi-squared of the fit.}
\begin{tabular}{|l|l|l|l|l|l|}

\hline
Sector  & Density Jump & M & $\alpha_1$ & $\alpha_2$ & $\chi^{2}/\nu$  \\ [0.5ex] 

\hline
$EP$$1$ & $ 2.55\pm 0.48$ &\multicolumn{1}{l|}{$2.3_{-0.35}^{+0.57}$} & $-0.96\pm 0.6 $ & $1.55\pm 0.3$ & $24.39/25$  \\
\hline
$EP$$2$ & $2.4 \pm 0.42 $ & $2.12_{-0.27}^{+0.41}$ & $-1.12 \pm 0.51$ & $1.95 \pm 0.44$ & $8.4/22$ \\
\hline
$EP$$3$ & $2.75 \pm 0.51$ & $2.58_{-0.43}^{+0.76}$& $-1.17 \pm 0.34 $ & $1.45 \pm 0.24$  & $42.2/40$ \\
\hline
$EP$$4$ & $3.01 \pm 0.64$ & $3.03_{-0.67}^{+1.62}$ & $ -0.64 \pm 0.36$ & $1.23 \pm 0.24$ & $38.2/40$ \\
\hline
$EP$$5$ & $2.76 \pm 0.52$ & $2.59_{-0.44}^{+0.79}$ & $-0.26 \pm 0.17$ & $1.29 \pm 0.22 $ & $41.85/46$\\
\hline
$EP$$6$ & $2.03 \pm0.33$ & $1.76_{-0.18}^{+0.24} $& $ -0.06 \pm 0.09$ & $ 1.45 \pm 0.2 $ & $27.50/31$ \\
\hline
$EP$$7$ & $2.57 \pm 0.53$ & $2.32_{-0.38}^{+0.64}$ & $0.05 \pm 0.17$ & $1.39 \pm 0.29 $ & $34.53/43$\\
\hline
$EP$$8$ & $2.25 \pm 0.52$ & $1.96_{-0.31}^{+0.47}$ & $0.46 \pm 0.12$ & $1.16 \pm 0.25 $ & $28.03/231$\\
\hline
$EP$$9$ & $2.17 \pm 0.45$ & $1.89_{-0.25}^{+0.37}$ & $0.34 \pm 0.17$ & $1.33 \pm 0.28 $ & $37.26/40$\\
\hline
$EP$$1-2$ & $2.58 \pm 0.43$ & $2.34_{-0.32}^{+0.5}$ & $-1.19 \pm 0.63$ & $1.72 \pm 0.31 $ & $15.61/19$\\
\hline
$EP$$3-6$ & $3.04 \pm 0.36$ & $3.09_{-0.43}^{+0.75}$ & $-0.47 \pm 0.11$ & $1.17 \pm 0.13 $ & $25.76/40$\\
\hline
$EP$$7-9$ & $1.84 \pm 0.56$ & $1.6_{-0.27}^{+0.39}$ & $-0.1 \pm 0.99$ & $1.65 \pm 0.30 $ & $9.81/19$\\
\hline
SE edge & $1.29 \pm 0.12$ & $1.19_{-0.05}^{+0.06}$ & $-0.9 \pm 0.26$ & $2.47 \pm 0.15 $ & $27.04/25$\\
\hline
\end{tabular}

\end{table*}

The surface brightness profiles can also be extracted by assuming elliptical geometry to better describe the geometry of the shock, following \cite{Ogrean_2014}. The ellipse chosen for this purpose has a major axis of 5 arcmin, a minor axis of 3.01 arcmin and an angle between the major axis and the right ascension axis of $330$ $^{\circ}$. The right-side panel of Fig. \ref{fig:ggm_3-all_sectors} shows the sectors (EP1-9) used to extract surface brightness profiles, where the white sectors represent the primary shock front, the green sector represents the SE edge and the circle filled in green is the region where BCG2 lies, and hence excluded before extracting the profiles.  

As described previously, once the surface brightness profiles are extracted, they are fitted with the broken power-law model to obtain the density profiles. Fig. \ref{fig:ep-1-9-sb} shows the surface brightness (in red) across each of these sectors. After subtracting the background (in solid black), these profiles are fitted with the broken power-law density model (in blue). Furthermore, the sectors are binned as EP1-2, EP3-6, and EP7-9 so that the sectors with the maximum gradient in intensity in the GGM as well as the highest density jumps are binned together. The surface brightness profiles across these binned sectors are seen in Fig. \ref {fig:ep_binned_-sb}. We see that the highest density jump is observed in the binned sector EP3-6, with $\rho = 3.04 \pm 0.36$ corresponding to a Mach number of $M = 3.09_{-0.43}^{+0.75}$. The density jumps, Mach numbers and the parameters used for the fitting for the individual and binned sectors, as well as the SE edge can be seen in table \ref{density_jump_sector_elliptical}. The results for the individual and binned sectors for spherical and elliptical geometry for individual and binned sectors are well in agreement with each other as seen in the comparison table \ref{tab:comp_table} in section \ref{section: appendix} (Appendix). 

\begin{figure*}
	\hbox{
	\includegraphics[width=0.5\linewidth]{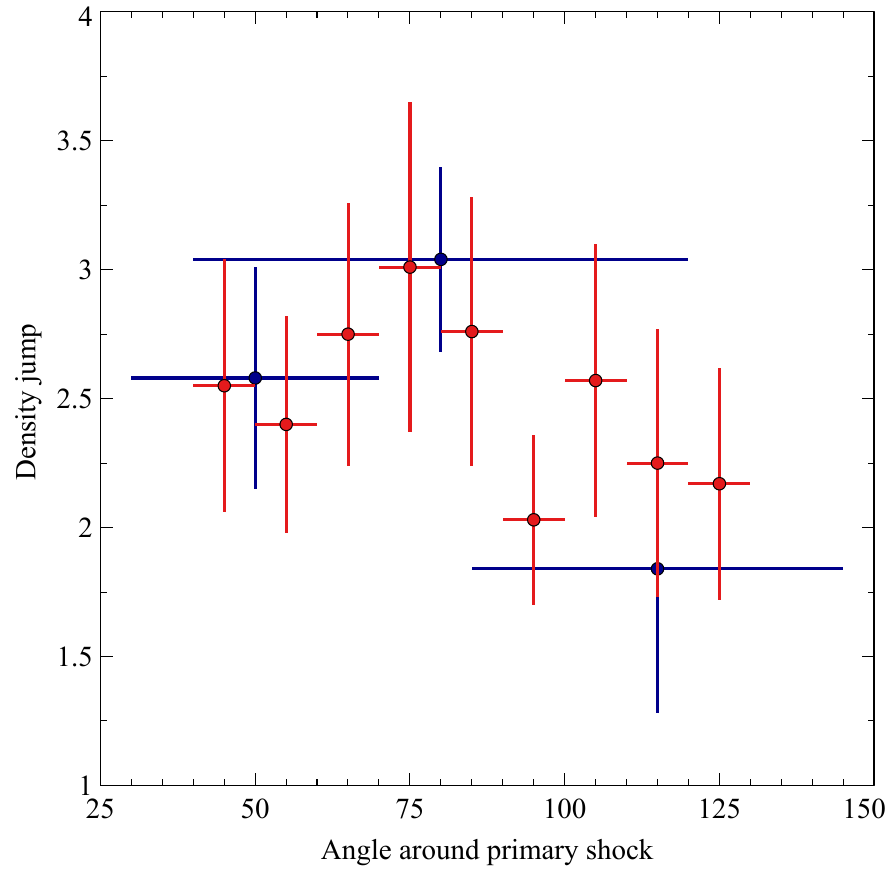}
	\includegraphics[width=0.5\linewidth]{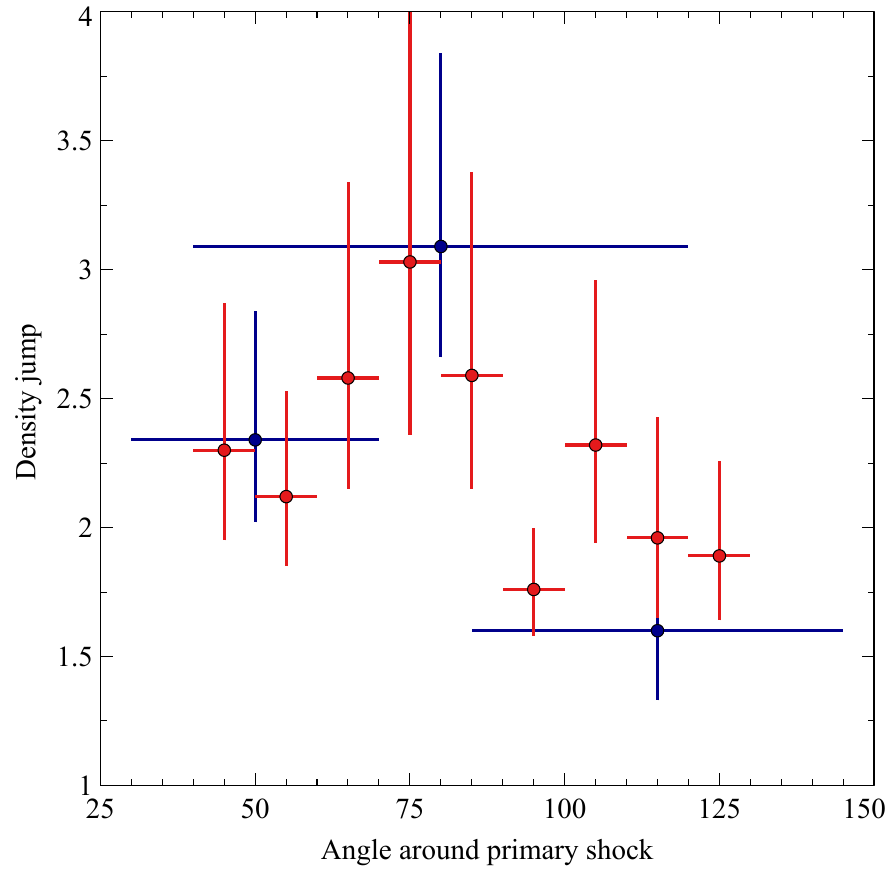}
	}
    \caption{ \textit{Left:} The cross-bars in red depict the density jump (elliptical geometry) across each of the sectors EP1 - EP9 and the ones in blue represent the density jumps in sectors EP1-2, EP3-6 and EP7-9 from left to right. The x-axis represents the angle of the sectors around the primary shock front, going from $45^{\circ}$ to $120^{\circ}$. \textit{Right:} The Mach number, determined from the density jump, is shown here in red for the sectors P1-9 and for sectors P1-2, P3-6, and P7-9 shown in blue, across the angles of the sectors around the primary shock front.    }
    \label{fig:dj_m_vs_angles_elliptical}
\end{figure*}

The density jump and Mach number obtained from the elliptical sectors EP1-9 along the Primary shock are plotted in Fig. \ref{fig:dj_m_vs_angles_elliptical}. The panel on the left shows the density jumps across the sectors plotted against the angle around the primary shock front going from $45^{\circ}$ to $125^{\circ}$. The red crosses indicate the density jump across each individual sector P1-9. The blue crosses represent the sectors binned as P1-2, P3-6, and P7-9. The panel on the right shows the Mach numbers derived from the corresponding density jumps using equation \ref{M_dens_jump}, also plotted against the angle around the primary shock front. The red crosses represent the Mach number derived for each of the sectors P1-9 and the blue crosses represent the elliptical binned sectors EP1-2, EP3-6, and EP7-9. 

Similar to Fig. \ref{fig:dj_m_vs_angles}, in both the plots, the observed trend is that the density jump and Mach number is highest at the center of the shock front where the GGM image shows the highest gradient.  On both sides of this center point, as the brightness of the GGM image decreases, the values of density jump and the Mach number taper off. The peak values of the density jump and Mach number are $3.04 \pm 0.36$ and $3.09_{-0.43}^{+0.75}$ respectively for the binned sector EP3-6, the brightest region in the GGM image. 

Based on the values of the Mach number for the sectors P3-6 and EP3-6 the primary shock in SPT J2031 is one of the strongest shocks, when compared with the Bullet Cluster with $ M = 3.0 \pm 0.4 $ \citep{Markevitch_2006_BC}, A2146 with $ M = 2.3 \pm 0.2  $ \citep{Russell_2010,Russell_2012,Russell_2022}, A665 with $ M = 3.0 \pm 0.6  $ \citep{Dasadia_2016}, A520 with $ M = 2.4_{-0.3}^{+0.4}  $ \citep{wang_2018}, and A98 with $M = 2.3 \pm 0.3 $ \citep{Sarkar_2022}.

\subsubsection{SE Edge}

\begin{figure*}
    \centering
    \hbox{
	\includegraphics[width=0.5\linewidth]{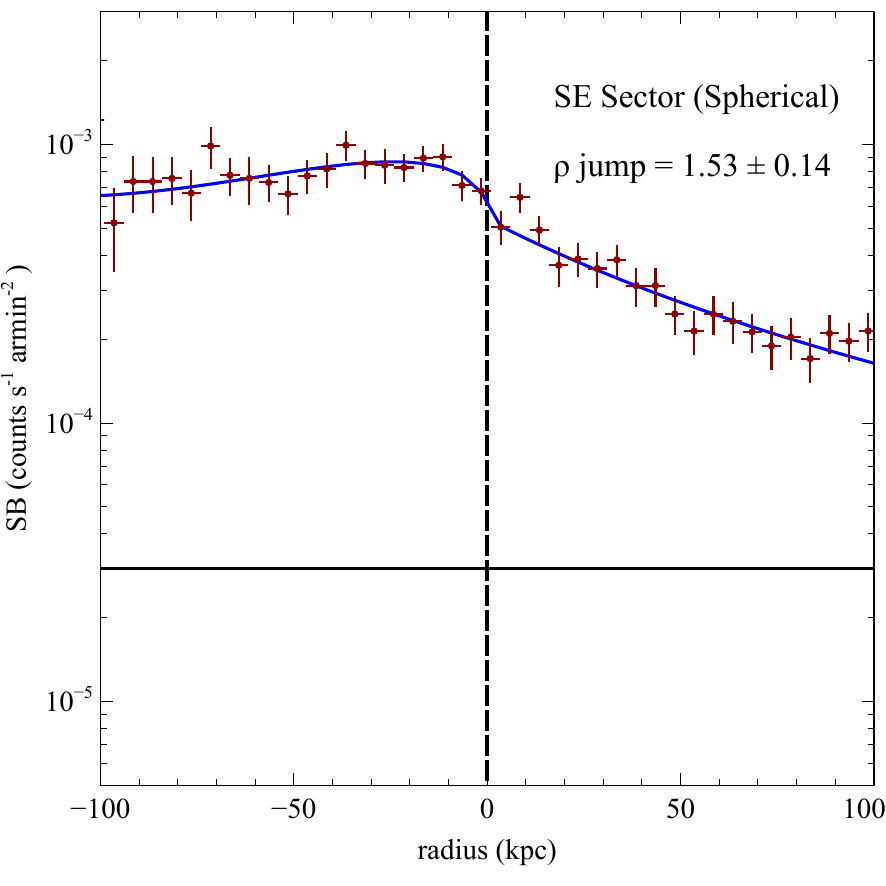}
	\includegraphics[width=0.5\linewidth]{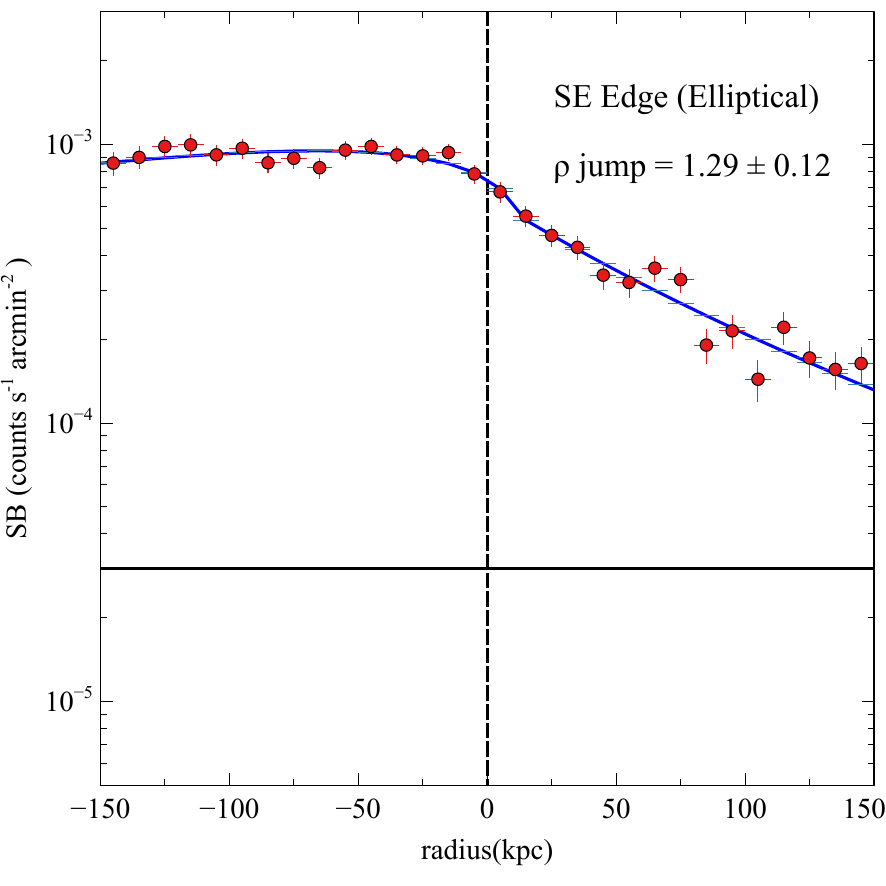}
	}
	\caption{Surface brightness profile extracted in the $0.5 - 2.5 keV$ energy band over the SE edge using spherical (left) and elliptical (right) geometry. The profiles were background subtracted and fitted with the broken power-law density model (solid blue line) to obtain density jump and Mach number. For the spherical geometry, the density jump obtained is $1.53 \pm 0.14$ which corresponds to a Mach number of $1.36_{-0.08}^{+0.09}$, whereas for the elliptical geometry, the density jump is $1.29 \pm 0.12$ corresponding to a Mach number of $1.19_{-0.05}^{+0.06}$. The dashed vertical line shows the shock location. }
    \label{se_temp}
\end{figure*}

Fig. \ref{se_temp} shows the surface brightness profile across the SE edge. The density jump across this edge is $1.53 \pm 0.14$ which corresponds to a Mach number of $1.36_{-0.08}^{+0.09}$. 
The analysis of the trend in the density jump and Mach number was not possible for the SE edge, as that edge is not spatially extended enough, and it does not have a high enough density jump or corresponding Mach number. 

\subsection{Spectral Analysis of the Shock fronts}

\begin{figure*}
	\centering
	\includegraphics[width=0.7\textwidth]{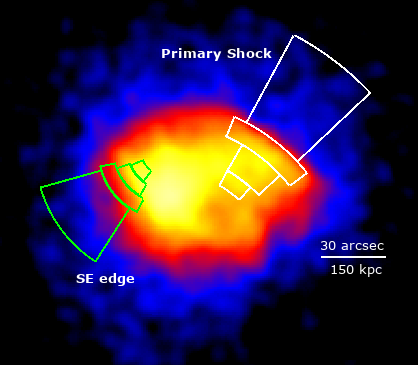}
	
    \caption{Exposure-corrected \textit{Chandra }image of SPT J2031 in the 0.5-2.5 keV energy range with the regions which were used to extract the temperature profiles across both the surface brightness edges. }
    \label{fig: exp_corr-temp_pro_regions}
\end{figure*}

\begin{figure*}
    \hbox{
	\includegraphics[width=0.5\linewidth]{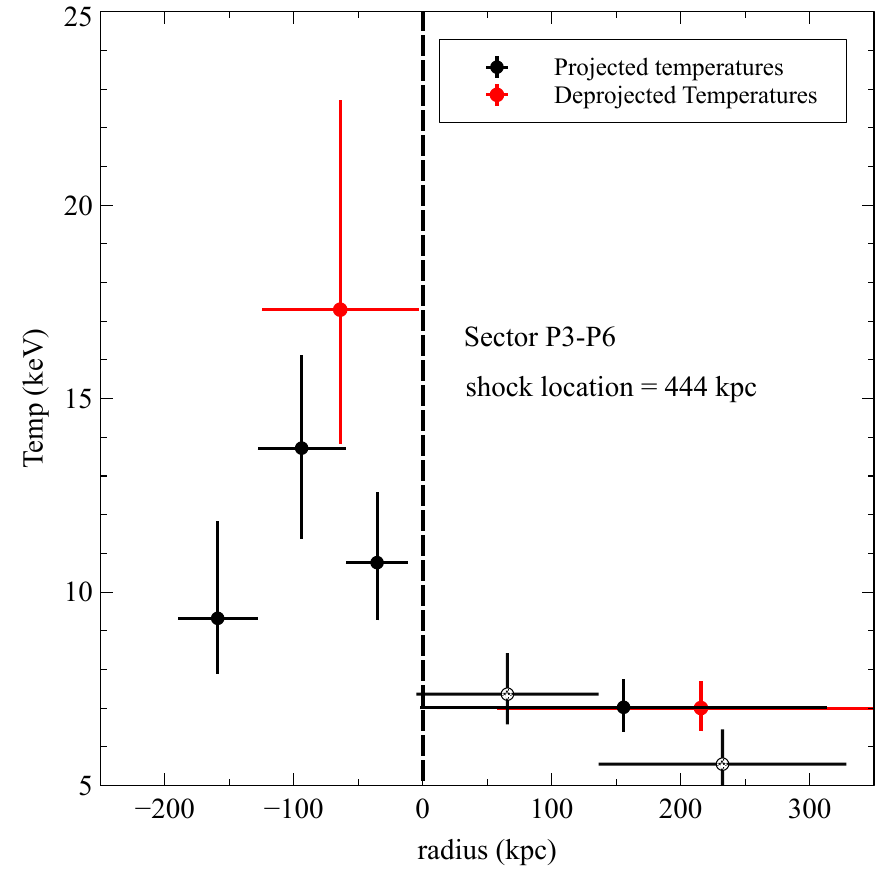}
   \includegraphics[width=0.5\linewidth]{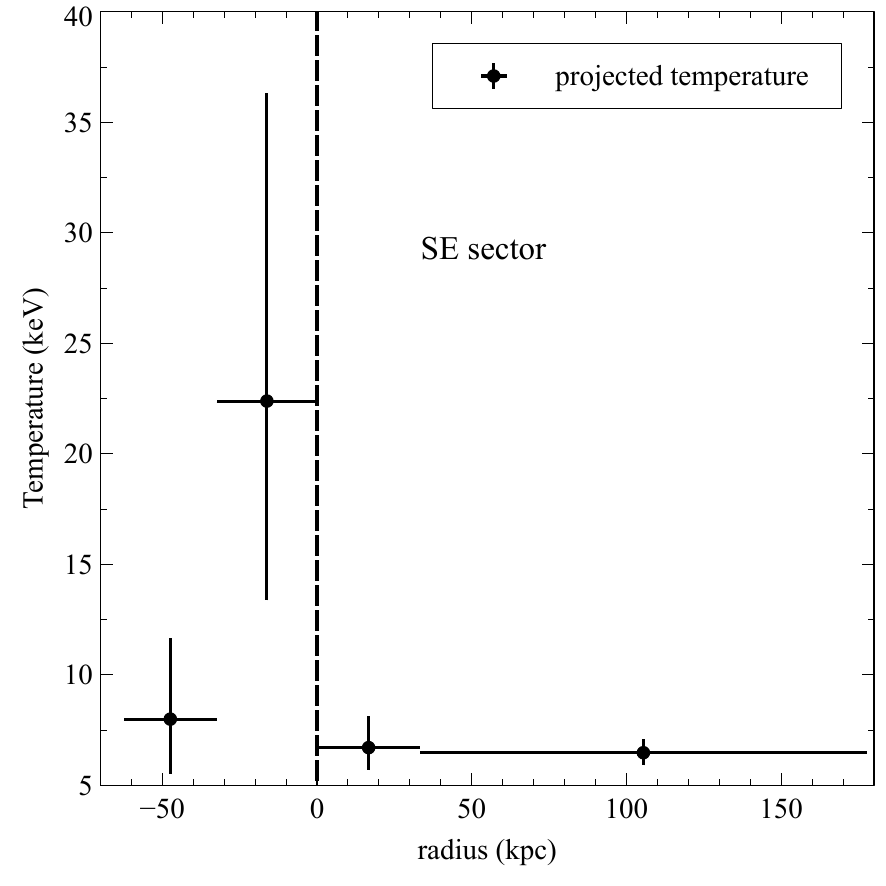}
   }
    
    \caption{\textit{Left:}The observed projected electron temperature profile(in black) over the primary shock front overlaid with the deprojected temperature profile (in red). The two data points with white circles in the pre-shock region represent narrower bins of width 150 kpc. We define the shock location to be at $r=0$ kpc. \textit{Right:}The figure shows the observed projected electron temperature profile across the SE edge. Again, we define the shock location to be at $r=0$ kpc.} 
    \label{fig:ps_se_temp_profile}
\end{figure*}

\begin{figure*}
    \hbox{
	\includegraphics[width=0.5\linewidth]{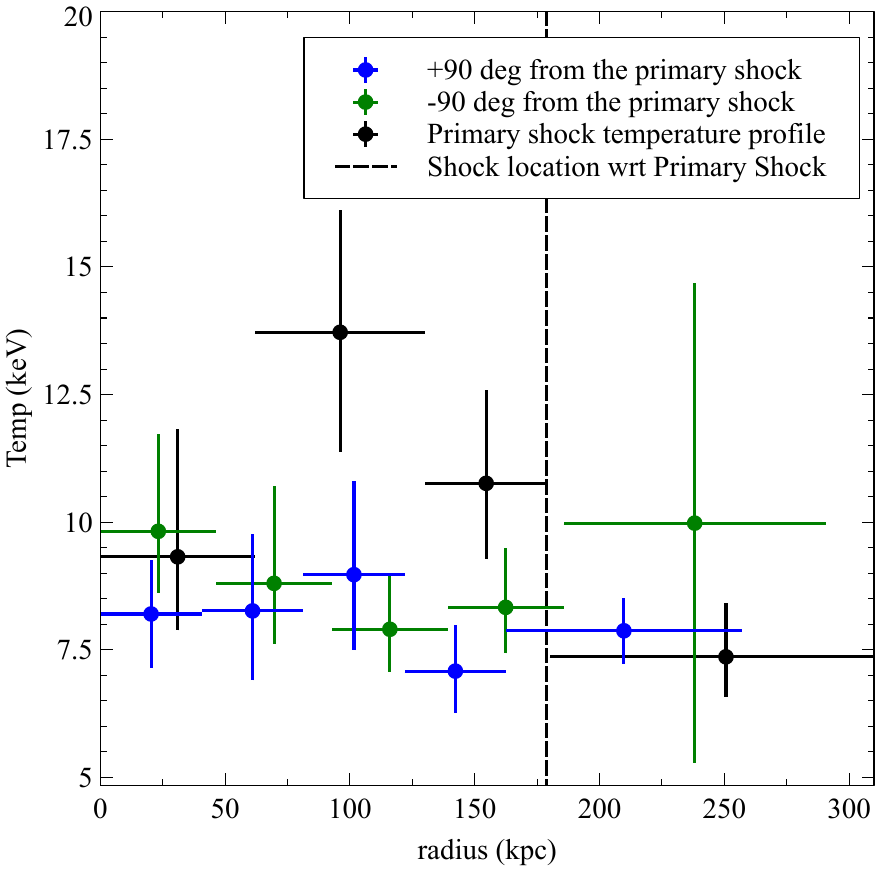}
   \includegraphics[width=0.5\linewidth]{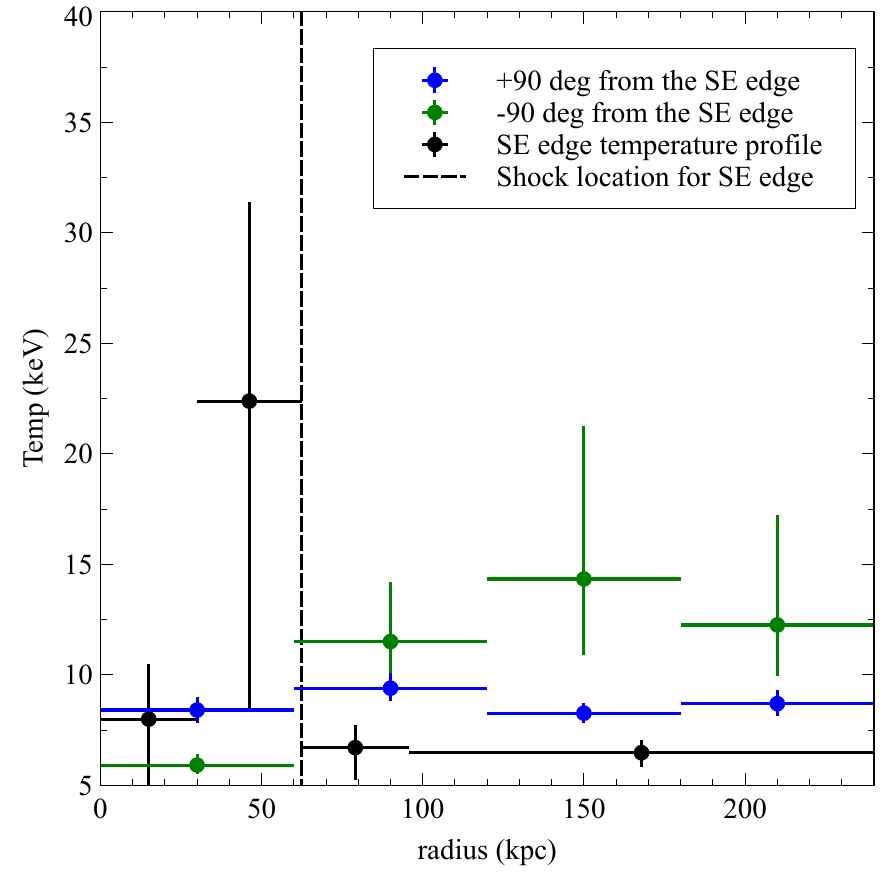}
   }
    
    \caption{\textit{Left:} Comparison of the temperature profile across the primary shock (in black) with the temperature profiles at +90 degrees  (in blue)  and at -90 degrees (in green) from the primary shock front. \textit{Right:} Comparison of the temperature profile across the SE edge (in black) with the temperature profiles at +90 degrees (in blue) and at -90 degrees (in green) from the SE edge}
    \label{fig:comparison_perpendicular}
\end{figure*}
The changes in temperature and density across the surface brightness edges can be observed more accurately by extracting radial profiles over the sectors shown in Fig. \ref{fig: exp_corr-temp_pro_regions}. The regions were selected so as to obtain the gas properties on both sides of each shock front. For the primary shock front, which has the higher density jump of the two, we extracted the temperature profile from the section of the shock with the highest density jump as determined in section \ref{sb_profiles} corresponding to the sector P3-6.  Using \textsl{specextract} in Sherpa, spectra were extracted from each of the regions for each of the ten observations. These spectra were then analysed in the energy range of $0.5 - 7.0$ keV. The background spectra used here were from the blank-sky backgrounds. The spectra for each region were then simultaneously fitted for all observations using Sherpa with the multiplicative PHABS(APEC) model, where the hydrogen column density is fixed at $n_H = 3.0 \times 10^{20}$ cm$^{-2}$ \citep{NH_Kalberla}, the solar abundance is $0.3$ Z$_\odot$, and the redshift is $0.34$ and chi-squared statistics were applied. 
The resulting projected temperature profile for Primary shock is shown in the left panel of  Fig. \ref{fig:ps_se_temp_profile}. We see a significant temperature jump from about $7.0_{-0.6}^{+0.7}$ keV to $13.8_{-1.8}^{+2.3}$ keV. The projected temperature profile for the SE edge is shown in the right panel of Fig. \ref{fig:ps_se_temp_profile}. We see a temperature jump from $6.48_{-0.57}^{+0.63}$ keV to $22.39_{-9.02}^{+13.92}$ keV.
The gas in the pre-shock region of the primary shock has a temperature of $7.0_{-0.6}^{+0.7} keV$. Along with the sharp increase in the surface brightness, there is an observed increase in the temperature in the post-shock region. The gas in this region has a temperature of $13.8_{-1.8}^{+2.3}$ keV. For the purpose of the fitting, the abundance is fixed at 0.3 $Z_{\odot}$. We performed necessary checks to compare the temperature profiles across both the shock fronts with the temperature profile of the cluster along perpendicular directions to both shocks.  Specifically, we extracted temperature profiles along the perpendicular directions of both shocks. Our analysis revealed no temperature jump in any of the perpendicular directions to both shocks. The left panel in Fig. \ref{fig:comparison_perpendicular} shows the temperature profile across the primary shock front (in black) compared with the temperature profiles at +90 $^{\circ}$ (in blue) and at -90 $^{\circ}$ (in green). Similarly, the plot on the right compares the temperature profile of the SE edge (in black) with the temperature profiles at +90 $^{\circ}$ in blue) and at -90 $^{\circ}$  (in green) from the SE edge. In both cases, it can be seen that the temperature profile remains mostly flat in regions that are not associated with either shock fronts.

Following \cite{Russell_2012}, we also obtained the values of the deprojected electron temperature using PROJCT in XSPEC. PROJCT is a deprojection routine that assumes spherical geometry for the cluster. This seems to be a reasonable assumption because the shocks in SPT J2031 appear to be approximately circular in the plane of the sky. The pre-shock electron temperature is $7.0_{-0.6}^{+0.7}$ keV and the post-shock deprojected electron temperature is $17.3_{-3.48}^{+5.41}$ keV. This allows us to calculate the Mach number using the deprojected temperature jump. The values for the deprojected temperature are plotted in the left panel Fig. \ref{fig:ps_se_temp_profile} (red crosses).

We use the following Rankine-Hugoniot equation for Mach number using the deprojected temperature jump:

\begin{equation}
    M = \left [ \frac{(\gamma +1)^2 ((\frac{T_2}{T_1})-1)}{2 \gamma (\gamma - 1)}\right] ^ \frac{1}{2}
    \label{M_temp_jump}
\end{equation}

\noindent where $T_2/T_1$ is the deprojected temperature jump. The Mach number from the temperature jump using the values obtained from the deprojected electron temperature after using this equation is $2.13_{-0.38}^{+0.4}$. The Mach number obtained using the temperature jump is lower than that obtained from the density jump, which is similar to what is observed in \cite{Russell_2012}.

\subsection{Electron-ion equilibrium}

The observation of shock fronts in cluster mergers provides valuable insights into the process of electron-ion equilibration within the ICM. 
Cluster merger shock fronts affect electrons and ions differently. As the shock front passes through the ICM, it heats the ions in the ICM gas immediately owing to their lesser thermal velocity and greater mass compared to the electrons \citep{Zuhone_2022}. This leads to a significant increase in ion temperature, not immediately observed in electron temperature, which eventually equilibrates with ion temperature. However, the mechanism of this shock heating remains debated \citep{wang_2018}. Presently the two models that can explain how the ICM gas is shock-heated are the adiabatic-collisional model and the instant shock-heating model.

The adiabatic-collisional model posits that protons and heavier ions experience dissipative heating, while electrons undergo adiabatic compression to a temperature much lower than that of ions. This scenario arises due to the differing velocities of electrons and ions relative to the shock front.  The ions move at a velocity lower than the shock, whereas the electrons, with a much lower mass compared to the ions, move at a much higher thermal velocity than the shock and, thus are adiabatically compressed  \citep{MV_review_2007}. The temperature of these adiabatically compressed electrons is given by:

\begin{equation}
    T_{e,2} = T_{e,1} \left(\frac{\rho_2}{\rho_1}\right)^{\gamma-1}
    \label{adiabatic_collision}
\end{equation} 
where $T_{e2}$ is the adiabatically compressed electron temperature; $T_{e1}$ is the pre-shock electron temperature, $\left(\frac{\rho_2}{\rho_1}\right)$ is the density jump, where $\rho_1$ and $\rho_2$ are the pre-shock and post-shock densities respectively, and $\gamma$ is the ratio of specific heats for a monoatomic gas. The electrons eventually undergo Coulomb collisions and attain thermal equilibrium with the ions over a timescale \citep{Sarazin_1986} given by: 

\begin{equation}
        \tau_{eq}(e,p) = 6.2 \times 10^8 yr  \left(\frac{n_e}{10^{-3}}\right)^{-1}  \left(\frac{T_e}{10^8 K}\right)^{3/2}
    \label{timescale}    
\end{equation}

\noindent where $n_e$ is the electron density and $T_e$ is the electron temperature. 

The instant shock heating model assumes that the intracluster medium (ICM) is a magnetized, collisionless plasma. This model has been proposed to explain observations of solar wind shocks, where electron and proton temperatures exhibit a jump on a linear scale of order several proton gyroradii, much smaller than their collisional mean free path \citep{MV_review_2007}. The coupling of particles with electric and magnetic fields results in interactions with dissipation of scale much shorter than the collision mean free path \citep{Russell_2012}. Hence, it is possible to find an electron-ion equilibration timescale shorter than the Coulomb timescale.

Following the conservation of the total kinetic energy density and assuming that the relation between the electron density ($n_e$) and ion density ($n_i$) is given by $n_e = 1.21 n_i$, the mean temperature of the gas 
 ($T_{gas}$) remains constant with time and is given by \citep{Zuhone_2022}:

\begin{equation}
    T_{gas} = \frac{(n_i T_i + n_e T_e)}{(n_i + n_e)} = \frac{T_i + 1.21 T_e}{2.21} 
    \label{mean_gas_temp}
\end{equation}

The rate at which the electron and ion temperatures equilibrate via Coulomb collisions is given by: 

\begin{equation}
    \frac{dT_e}{dt} = \frac{T_{i}-T_{e}}{t_{eq}}
    \label{coulomb_collision_rate}
\end{equation}

Rearranging this equation,

\begin{equation}
    \frac{t_{eq}}{T_{i}-T_{e}} dT_{e} = dT
    \label{dT_equation}
\end{equation}

Integrating Equation \ref{dT_equation} one can obtain the model electron temperature analytically (see \cite{Ettori-fabian}). The emissivity-weighted electron temperature profile is projected along the line of sight by:

\begin{equation}
    <T> = \int_{b^2}^{\infty} \frac{\epsilon(r) T_e(r) dr^2}{\sqrt{r^2-b^2}} \bigg/ \int_{b^2}^{\infty} \frac{\epsilon(r) dr^2}{\sqrt{r^2-b^2}} 
\end{equation}

where $\epsilon (r)$ is the emissivity at radius $r$ and $b$ is the distance from the shock front \citep{Sarkar_2022}.

\begin{figure*}
	\hbox{
	\includegraphics[width=0.5\linewidth]{images/p1234_sb_plot.pdf}
	\includegraphics[width=0.5\linewidth]{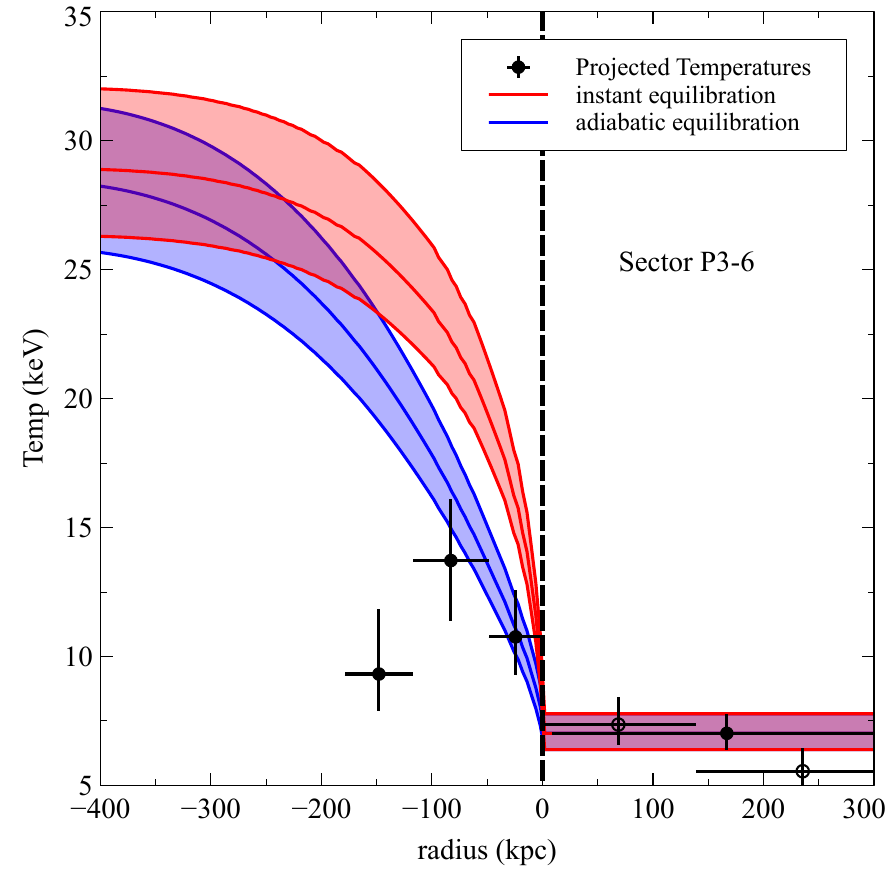}
	}
    \caption{\textit{Left:} Surface brightness profiles extracted over sectors $P$ $3-6$ in the $0.5 - 2.5$ keV energy band with BCG2 excluded. The profiles are background subtracted and have been fitted with the broken power-law gas density model (in solid blue). The density jump observed over this sector is $3.16 \pm 0.34$ and the Mach number resulting from this density jump is $3.36_{-0.48}^{+0.87}$. 
    \textit{Right:} Projected electron temperatures (in black) observed across the sectors $P$ $3-6$ of the primary shock front compared with the overlaid adiabatic-collisional (in blue) and instant heating models (in red) projected (upto $1 \sigma$ error bands) along the line of sight for electron-ion equilibration. The pre-shock region shows the temperature value for a 300 kpc wide bin(in solid black) as well as two smaller bins of 150 kpc width (white-filled circles) each.  The post-shock temperature for the primary shock front seemingly favours the Adiabatic-collisional model over the instant heating model. }
    \label{fig:p3-6-sb-models}
\end{figure*}




Although we cannot measure the temperature of the ions, it is possible to measure the jump in the gas density across the shock front (which we have done in section \ref{sb_profiles}) which can be used to calculate the post-shock equilibrium temperature for the electrons and ions using the Rankine-Hugoniot jump conditions from the pre-shock temperature \citep{Landau_Lifshitz_1959}. The temperature jump can be obtained using equation \ref{temp_jemp} \citep{MV_review_2007}.

\begin{equation}
    \frac{T_{2}}{T_{1}} = \frac{\zeta-\rho_{1}/\rho_{2}}{\zeta-\rho_{2}/\rho_{1}} 
    \label{temp_jemp}
\end{equation}

\noindent where we have assumed that $\gamma = 5/3 $, the adiabatic index for monoatomic gas, and $\zeta \equiv (\gamma + 1)/(\gamma - 1)$, and the indices 1 and 2 denote the pre-shock and post-shock quantities respectively. 


We did this for both the instant heating model and the collisional equilibration model as can be seen in Fig. \ref{fig:p3-6-sb-models}.

The bow shock in the Bullet Cluster provided the first-ever opportunity to determine the timescale of the electron-ion equilibration in the magnetized ICM \citep{Markevitch_2006_BC}. The two models of heating were compared with the observed electron temperature profile. \cite{Markevitch_2006_BC} shows tantalising evidence supporting the instant equilibration model at a 95$\%$ significance, as the post-shock temperature in the cluster is very high($\sim 20-40$ keV) and hence very difficult to constrain.

The post-shock temperatures for SPT-CLJ 2031-4037 are much lower than the Bullet Cluster ($\sim 13-15$ keV) and hence can be constrained much better. Additionally, the Mach number is sufficiently high to obtain the separation between the adiabatic and instant equilibration models.

The right-hand panel of Fig. \ref{fig:p3-6-sb-models} shows how the electron temperatures observed across the primary shock front compare with the collisional and instant heating models projected along the line of sight. The observed post-shock temperature for the primary shock front seems to favour the Collisional model over the instant heating model. 

The models were generated by assuming spherical geometry. The cluster, which is assumed to be spherical is divided into 1000 shells of uniform radii and thereby has the same volume dV. 
The electron temperature along the line of sight was obtained for each shell and the corresponding emissivity is also obtained based on the density profile. The projected models were determined from the emissivity-weighted electron temperature. Using eq \ref{timescale}. the electron-ion equilibration timescale for Coulomb collisions is calculated to be $0.2$ Gyr.

In the case of the Bullet Cluster, \cite{Markevitch_2006_BC} found that the observed temperature profile supports the instant equilibration model, suggesting that electrons at the shock front were heated on a timescale faster than the Coulomb collisional timescale. However, the post-shock temperature in the Bullet cluster is $\sim 20 - 40$ keV, which is much higher than the energy pass band of \textit{Chandra}, thus making it difficult to constrain. The post-shock electron temperature in SPT J2031 is lower than that of the Bullet Cluster, making the measurements of the post-shock temperature more accurate. In contrast, an analysis of the shock in the Bullet Cluster by ALMA and ACA \citep{Mascolo_bc_sz} found that the assumption of an adiabatic temperature jump in the electron temperature results in the best agreement between results of Sunyaev-Zeldovich and X-ray measurements. 

For the merger shock front in A2146, \cite{Russell_2012} found that the the temperature profile across the bow shock is consistent with the collisional equilibration model, whereas the upstream shock favours the instant equilibrational model. However, the uncertainty in the measurement for the upstream shock was higher because of its lower Mach number and hence was not determined to be the definite conclusion. Subsequently, with deeper 2 Ms \textit{Chandra} observations of A2146, \cite{Russell_2022} found that both the shock fronts support the collisional equilibration model.    
Our results for the primary shock in SPT J2031 agree with \cite{Russell_2012,Russell_2022} in that the observed post-shock electron temperature favours the Collisional equilibration model. 

Analysis of the merger shock front in A520 \citep{wang_2018} found that the post-shock electron temperature was higher than expected from a situation where the electrons undergo adiabatic compression followed by Coulomb collisions. Hence, like the Bullet Cluster, the electron temperature profile in A520 also supports the instant equilibration model with a confidence level of 95$\%$. 

A similar comparison of the post-shock electron temperature in the merger shock of A98 \citep{Sarkar_2022} with the Collisional and instant equilibrational model showed that the observed post-shock electron temperature favors the instant equilibration model, however, the large uncertainties in the temperature indicate that the Collisional model can not be ruled out.

The pre-shock sound speed, derived from the equations $c_s = \sqrt{\gamma k_B T / m_H \mu}$ is $ (1.3 \pm 0.06) \times 10^3$ km s$^{-1}$. The shock speed, obtained by multiplying the Mach number (from the density jump, $M = 3.23_{-0.56}^{+0.89}$) and the sound speed is $v_{shock} \sim (4.4_{-0.16}^{+0.27}) \times 10^3$ km s$^{-1}$. The post-shock velocity $v_{ps}$ for the primary shock front is $1414$ km/s obtained by dividing the shock speed by the density jump. 




\section{Conclusions}
We conducted a comprehensive analysis of our newly acquired  deep ($256$ ks) \textit{Chandra} observations of the merging system SPT J2031 and obtained the following results:
\begin{itemize}
  \item SPT J2031 exhibits merger geometry, as suggested by an offset between the brightest X-ray peaks in the exposure-corrected image from the \textit{Chandra} observations and the two Brightest Cluster Galaxies in the HST optical image. 
  \item We have utilised the GGM filtering technique to identify two sharp surface brightness edges in SPT J2031, the primary shock front and the southeastern edge. 
  \item We extracted surface brightness profiles (assuming spherical and elliptical geometries) across both the edges identified in the GGM image and fitted them with the broken power-law model to find the density jump across the shock front. The sharp edge in the northwest direction is the primary shock with a density jump $ \rho = 3.16 \pm 0.34$ corresponding to a Mach number of $3.36_{-0.48}^{+0.87}$ for spherical geometry and a density jump of $ \rho = 3.04 \pm 0.36$ corresponding to a Mach number of $3.09_{-0.43}^{+0.75}$ for elliptical geometry. 
  \item Due to the high Mach number obtained from the density jump in the primary shock front, we were able to compare the observed electron temperature profile of the primary shock with the collisional equilibration model and the instant shock heating model. We found that the post-shock electron temperature is lower than the temperature predicted for the instant shock heating model and favours the collisional equilibrational model. These findings are similar to the result in \citet{Russell_2012, Russell_2022}.  However, we cannot completely rule out the instant heating model.
  \item The other surface brightness edge, the SE edge is observed in the southeastern direction and also appears to be a shock front. It has a  density jump  $ \rho = 1.53 \pm 0.14 $ corresponding to a Mach number $M = 1.36_{-0.08}^{+0.09}$ for spherical geometry and a  density jump  $ \rho = 1.29 \pm 0.12 $ corresponding to a Mach number $M = 1.19_{-0.05}^{+0.06}$. Since the Mach number $M < 2$, we were not able to achieve enough separation between the two projected models of heating to compare the observed electron temperature profile.

  \item We plotted the density jump and Mach number of the primary shock as a function of the angle around the shock front and found that the density jump, and subsequently the Mach number peak at the center of the shock front, where the gradient in the GGM image is maximum. Both the density jump and the Mach number taper off with a change in angle on both sides of this center point. 
 
\end{itemize}

\section*{Acknowledgements}
We gratefully acknowledge the valuable feedback provided by the referee, which enhanced the quality and clarity of this manuscript. We acknowledge support from Chandra grant GO1-22121X. We thank Florian Hofmann for the helpful discussions.  This work is based
on observations obtained with the Chandra observatory, a NASA
mission.

\section*{Data Availability}
This paper employs a list of Chandra datasets, obtained by the Chandra X-ray Observatory, contained in the Chandra Data Collection (CDC) \dataset[DOI: https://doi.org/10.25574/cdc.229] 
{https://doi.org/10.25574/cdc.229}.

\software{CIAO \citep{CIAO_software}, 
        XSPEC \citep{Xspec_1996}, 
        Proffit \citep{proffit}}
\bibliography{ref}{}
\bibliographystyle{aasjournal}

\newpage
\section*{Appendix} \label{section: appendix}

As a part of our analysis, we extracted surface brightness profiles assuming spherical geometry without excluding the region containing BCG2 for completeness, and to check if BCG2 indeed affected the brightness. Fig. \ref{fig:ggm_3-all_sectors} compares the sectors used for this analysis with the sectors used for spherical and elliptical geometry where BCG2 has been excluded. Table \ref{tab:comp_table} compares the values for density jumps and corresponding Mach numbers obtained for all three cases. We find that for most regions, whether individual or binned, the results are well in agreement with each other.

\begin{figure}
	\includegraphics[width=1.0
	\linewidth]{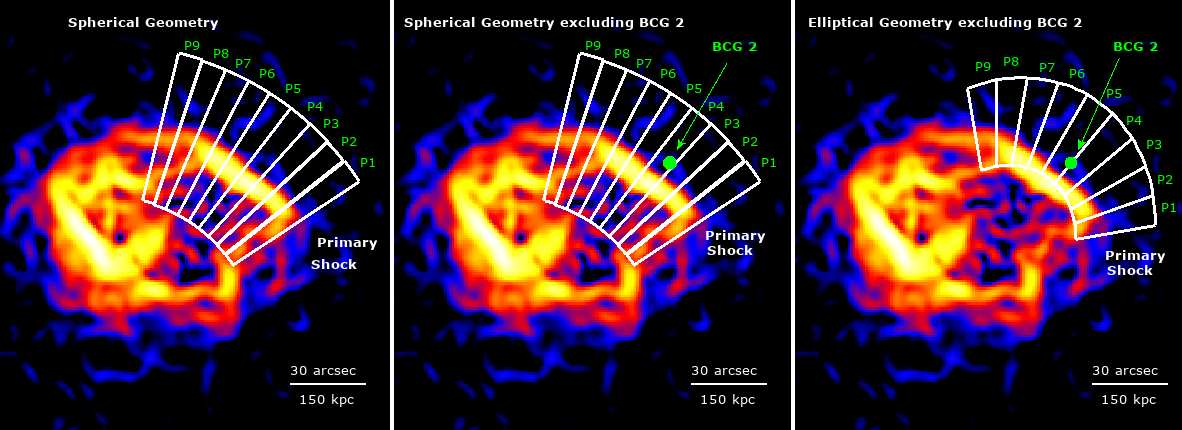}
	
    \caption{GGM image  }
    \label{fig:ggm_3-all_sectors-geometry}
\end{figure}

\begin{table}
\label{tab:comp_table}
 \caption{Table of comparison of density jumps and corresponding mach numbers  }
\begin{tabular}{|llllll|lll|l}
\cline{1-9}
\multicolumn{6}{|c|}{\textbf{Spherical Geometry}} &
  \multicolumn{3}{c|}{\textbf{Elliptical Geometry}} &
   \\ \cline{1-9}
\multicolumn{3}{|c|}{\textbf{W/ BCG 2}} &
  \multicolumn{3}{c|}{\textbf{W/O BCG 2}} &
  \multicolumn{3}{c|}{\textbf{W/O BCG 2}} &
   \\ \cline{1-9}
\multicolumn{1}{|l|}{Sector} &
  \multicolumn{1}{l|}{Density Jump} &
  \multicolumn{1}{l|}{Mach Number} &
  \multicolumn{1}{l|}{Sector} &
  \multicolumn{1}{l|}{Density Jump} &
  Mach Number &
  \multicolumn{1}{l|}{Sector} &
  \multicolumn{1}{l|}{Density Jump} &
  Mach Number &
   \\ \cline{1-9}
\multicolumn{1}{|l|}{$P$$1$} &
  \multicolumn{1}{l|}{$2.21 \pm 0.49$} &
  \multicolumn{1}{l|}{$1.92_{-0.38}^{+0.53}$} &
  \multicolumn{1}{l|}{$P$$1$} &
  \multicolumn{1}{l|}{$2.21 \pm 0.55$} &
  \multicolumn{1}{l|}{$1.92_{-0.29}^{+0.43}$}&
  \multicolumn{1}{l|}{$EP$$1$} &
  \multicolumn{1}{l|}{$2.55 \pm 0.49$} &
  \multicolumn{1}{l|}{$2.3_{-0.35}^{+0.57}$}&
   \\ \cline{1-9}
\multicolumn{1}{|l|}{$P$$2$} &
  \multicolumn{1}{l|}{$2.8 \pm 0.55$} &
  \multicolumn{1}{l|}{$2.65_{-0.83}^{+1.39}$} &
  \multicolumn{1}{l|}{$P$$2$} &
  \multicolumn{1}{l|}{$2.8 \pm 0.52$} &
  \multicolumn{1}{l|}{$2.65_{-0.83}^{+1.39}$} &
  \multicolumn{1}{l|}{$EP$$2$} &
  \multicolumn{1}{l|}{$2.4 \pm 0.42$} &
   \multicolumn{1}{l|}{$2.12_{-0.27}^{+0.41}$}&
   \\ \cline{1-9}
\multicolumn{1}{|l|}{$P$$3$} &
  \multicolumn{1}{l|}{$3.13 \pm 0.42$} &
  \multicolumn{1}{l|}{$3.28_{-0.71}^{+1.38}$} &
  \multicolumn{1}{l|}{$P$$3$} &
  \multicolumn{1}{l|}{$3.15 \pm 0.7$} &
   \multicolumn{1}{l|}{$3.35_{-0.85}^{+1.95}$}&
  \multicolumn{1}{l|}{$EP$$3$} &
  \multicolumn{1}{l|}{$2.75 \pm 0.51$} &
   \multicolumn{1}{l|}{$2.58_{-0.43}^{+0.76}$}&
   \\ \cline{1-9}
\multicolumn{1}{|l|}{$P$$4$} &
  \multicolumn{1}{l|}{$2.86 \pm 0.43$} &
  \multicolumn{1}{l|}{$2.74_{-0.55}^{+0.85}$} &
  \multicolumn{1}{l|}{$P$$4$} &
  \multicolumn{1}{l|}{$2.99 \pm 0.49$} &
   \multicolumn{1}{l|}{$2.99_{-0.52}^{+1.03}$}&
  \multicolumn{1}{l|}{$EP$$4$} &
  \multicolumn{1}{l|}{$3.01 \pm 0.64$} &
   \multicolumn{1}{l|}{$3.03_{-0.67}^{+1.62}$}&
   \\ \cline{1-9}
\multicolumn{1}{|l|}{$P$$5$} &
  \multicolumn{1}{l|}{$2.78 \pm 0.76$} &
  \multicolumn{1}{l|}{$2.61_{-0.79}^{+1.39}$} &
  \multicolumn{1}{l|}{$P$$5$} &
  \multicolumn{1}{l|}{$2.78 \pm 0.77$} &
   \multicolumn{1}{l|}{$2.63_{-0.62}^{+1.41}$}&
  \multicolumn{1}{l|}{$EP$$5$} &
  \multicolumn{1}{l|}{$2.76 \pm 0.52$} &
    \multicolumn{1}{l|}{$2.59_{-0.44}^{+0.79}$}&
   \\ \cline{1-9}
\multicolumn{1}{|l|}{$P$$6$} &
  \multicolumn{1}{l|}{$2.86 \pm0.57$} &
  \multicolumn{1}{l|}{$2.74_{-0.73}^{+1.50}$} &
  \multicolumn{1}{l|}{$P$$6$} &
  \multicolumn{1}{l|}{$3.13 \pm 0.6$} &
  \multicolumn{1}{l|}{$3.3_{-0.73}^{+1.91}$} &
  \multicolumn{1}{l|}{$EP$$6$} &
  \multicolumn{1}{l|}{$2.03\pm 0.33$} &
   \multicolumn{1}{l|}{$1.76_{-0.18}^{+0.24}$} &
   \\ \cline{1-9}
\multicolumn{1}{|l|}{$P$$7$} &
  \multicolumn{1}{l|}{$2.27 \pm 0.49$} &
  \multicolumn{1}{l|}{$1.98_{-0.4}^{+0.54}$} &
  \multicolumn{1}{l|}{$P$$7$} &
  \multicolumn{1}{l|}{$2.23 \pm 0.49$} &
  \multicolumn{1}{l|}{$1.94_{-0.28}^{+0.43}$} &
  \multicolumn{1}{l|}{$EP$$7$} &
  \multicolumn{1}{l|}{$2.57 \pm 0.53$} &
   \multicolumn{1}{l|}{$2.32_{-0.38}^{+0.64}$} &
   \\ \cline{1-9}
\multicolumn{1}{|l|}{$P$$8$} &
  \multicolumn{1}{l|}{$1.77 \pm 0.28$} &
  \multicolumn{1}{l|}{$1.54_{-0.19}^{+0.21}$} &
  \multicolumn{1}{l|}{$P$$8$} &
  \multicolumn{1}{l|}{$1.77 \pm 0.29$} &
   \multicolumn{1}{l|}{$1.54_{-0.14}^{+0.18}$}&
  \multicolumn{1}{l|}{$EP$$8$} &
  \multicolumn{1}{l|}{$2.25 \pm 0.52$} &
   \multicolumn{1}{l|}{$1.96_{-0.31}^{+0.47}$} &
   \\ \cline{1-9}
\multicolumn{1}{|l|}{$P$$9$} &
  \multicolumn{1}{l|}{$1.42 \pm 0.22$} &
  \multicolumn{1}{l|}{$1.28_{-0.14}^{+0.14}$} &
  \multicolumn{1}{l|}{$P$$9$} &
  \multicolumn{1}{l|}{$1.4 \pm 0.21$} &
   \multicolumn{1}{l|}{$1.27_{-0.09}^{+0.12}$}&
  \multicolumn{1}{l|}{$EP$$9$} &
  \multicolumn{1}{l|}{$2.17 \pm 0.45$} &
   \multicolumn{1}{l|}{$1.89_{-0.25}^{+0.37}$} &
   \\ \cline{1-9}
\multicolumn{1}{|l|}{$P$$1-2$} &
  \multicolumn{1}{l|}{$2.53 \pm 0.42$} &
  \multicolumn{1}{l|}{$2.27_{-0.31}^{+0.39}$} &
  \multicolumn{1}{l|}{$P$$1-2$} &
  \multicolumn{1}{l|}{$2.47 \pm 0.34$} &
   \multicolumn{1}{l|}{$2.2_{-0.23}^{+0.35}$}&
  \multicolumn{1}{l|}{$EP$$1-2$} &
  \multicolumn{1}{l|}{$2.58 \pm 0.43$} &
  \multicolumn{1}{l|}{$2.34_{-0.32}^{+0.5}$}  &
   \\ \cline{1-9}
\multicolumn{1}{|l|}{$P$$3-6$} &
  \multicolumn{1}{l|}{$3.11 \pm 0.32$} &
  \multicolumn{1}{l|}{$3.23_{-0.56}^{+0.89}$} &
  \multicolumn{1}{l|}{$P$$3-6$} &
  \multicolumn{1}{l|}{$3.16 \pm 0.34$} &
   \multicolumn{1}{l|}{$3.36_{-0.48}^{+0.87}$}&
  \multicolumn{1}{l|}{$EP$$3-6$} &
  \multicolumn{1}{l|}{$3.04 \pm 0.36$} &
 \multicolumn{1}{l|}{$3.09_{-0.43}^{+0.75}$}   &
   \\ \cline{1-9}
\multicolumn{1}{|l|}{$P$$7-9$} &
  \multicolumn{1}{l|}{$1.90 \pm 0.18$} &
  \multicolumn{1}{l|}{$1.64_{-0.13}^{+0.15}$} &
  \multicolumn{1}{l|}{$P$$7-9$} &
  \multicolumn{1}{l|}{$1.89 \pm 0.19$} &
   \multicolumn{1}{l|}{$1.64_{-0.09}^{+0.12}$}&
  \multicolumn{1}{l|}{$EP$$7-9$} &
  \multicolumn{1}{l|}{$1.84 \pm 0.56$} &
 \multicolumn{1}{l|}{$1.6_{-0.27}^{+0.39}$}   &
   \\ \cline{1-9}
  
\end{tabular}
\end{table}



\end{document}